\documentclass[aps,prd,onecolumn,groupedaddress,showpacs,nofootinbib,amssymb]{revtex4}
\usepackage[dvips]{graphicx}
\usepackage{amssymb}
\usepackage{amsmath}
\usepackage{graphicx,color}
\usepackage{amsfonts}
\usepackage{bm}
\usepackage{cancel}
\usepackage{comment}
\usepackage{hyperref}
\usepackage{ulem}

\newcommand{\e}{\mathrm{e}}

\allowdisplaybreaks[4]

\begin{document}

\tolerance=5000

\preprint{<KEK-TH-2629, KEK-Cosmo-0346>}
\title{
$F(Q)$ gravity with Gauss-Bonnet corrections: \\
from early-time inflation to late-time acceleration 
}

\author{Shin'ichi~Nojiri$^{1,2}$}\email{nojiri@gravity.phys.nagoya-u.ac.jp}
\author{S.~D.~Odintsov$^{3,4}$}\email{odintsov@ice.csic.es}

\affiliation{ $^{1)}$ Theory Center, IPNS, KEK, 1-1 Oho, Tsukuba, Ibaraki 305-0801, Japan \\
$^{2)}$ Kobayashi-Maskawa Institute for the Origin of Particles and the Universe, Nagoya University, Nagoya 464-8602, Japan \\
$^{3)}$ Institute of Space Sciences (ICE, CSIC) C. Can Magrans s/n, 08193 Barcelona, Spain \\
$^{4)}$ ICREA, Passeig Lluis Companys, 23, 08010 Barcelona, Spain
}

\begin{abstract}

We show that in the $f(Q)$ gravity with a non-metricity scalar $Q$, the curvatures in Einstein's gravity, that is, the Riemann curvature 
constructed from the standard Levi-Civita connection, could not be excluded or naturally appear. 
The first observation is that even in $f(Q)$ gravity, the conservation of the matter energy-momentum tensor is not described by 
the covariant derivatives in the non-metricity gravity but that is given by the Levi-Civita connection. 
The commutator of the covariant derivatives in Einstein's gravity inevitably induces the Riemann curvature. 
There is no symmetry nor principle which prohibits the Riemann curvature in non-metricity gravity. 
Based on this observation, we propose and investigate $f\left(Q, \mathcal{G} \right)$ gravity with the Gauss-Bonnet invariant $\mathcal{G}$ and its generalisations. 
We show how  $f\left(Q, \mathcal{G} \right)$ models realising any given the Friedmann-Lema\^{i}tre-Robertson-Walker (FLRW) spacetime can be reconstructed. 
We apply the reconstruction formalism to cosmology. 
Explicitly, the gravity models which realise slow roll or constant roll inflation, dark energy epoch as well as the unification of the inflation and dark energy are found. 
The dynamical autonomous system and the gravitational wave in the theory under investigation are discussed. 
It is found the condition that the de Sitter spacetime becomes the (stable) fixed point of the system. 

\end{abstract}

\maketitle

\section{Introduction\label{SecI}}

Recently different versions of modified gravity theories have been very actively studied (see reviews \cite{Capozziello:2011et, Faraoni:2010pgm, Cai:2015emx, Nojiri:2010wj, Nojiri:2017ncd}). 
Indeed, there appear more and more indications that there should be a more fundamental gravity theory beyond Einstein's general relativity, which so far
is a rather successful candidate for classical gravity. 
In the framework of general relativity, we find many unsolved problems like a dark energy problem and H0-tension. 
Especially when we consider quantum aspects of gravity, the information loss problem is caused by a black hole. So far we have not succeeded 
in constructing a realistic theory of renormalisable quantum gravity. 
Modified gravities are not the final theory but they may provide useful hints for the final fundamental theory of gravity. 

One important question is whether gravity should be formulated in terms of the Riemann curvatures, that is, the curvature given by the Levi-Civita connection. 
Gravity with the torsion $T$ \cite{Hehl:1976kj, Hayashi:1979qx, Bengochea:2008gz, Linder:2010py} instead of the Riemann curvature and its recent variant 
$f(T)$ gravity \cite{Li:2010cg, Bamba:2010wb, Chen:2010va, Dent:2010nbw, Cai:2015emx} have been actively studied.

More recently, the theories based on non-metricity tensors have been 
widely discussed \cite{Nester:1998mp, BeltranJimenez:2018vdo, Runkla:2018xrv, Capozziello:2022tvv}. 
In this theory, a fundamental geometrical quantity is the non-metricity scalar $Q$ and the connection is another independent variable. 
By imposing conditions that the Riemann tensor and torsion tensor vanish, the connection is written by the four scalar 
fields \cite{Blixt:2023kyr, BeltranJimenez:2022azb, Adak:2018vzk, Tomonari:2023wcs}. 
For the general covariance, it is often chosen the gauge condition that the connection vanishes, which is called the coincident gauge. 
When the action is linear in $Q$, the theory is equivalent to Einstein's general relativity. 
This is because the difference between $Q$ and the scalar curvature in Einstein's gravity is the total derivative. 

One may consider an analogue of $f(R)$ gravity or $f(T)$ gravity for the symmetric teleparallel theory, that is, $f(Q)$ gravity, where $f(Q)$ 
is a Lagrangian density and a function of $Q$. 
Especially, the studies on the dynamical degrees of freedom (DOF) have been thoroughly done 
\cite{Hu:2022anq, DAmbrosio:2023asf, Heisenberg:2023lru, Paliathanasis:2023pqp, Dimakis:2021gby, Hu:2023gui}.
The problem has not been completely solved although the only propagating mode in the flat background is a graviton~\cite{Capozziello:2024vix}, 
whose situation has not changed from Einstein's gravity and is similar to $f(T)$ gravity \cite{Bamba:2013ooa}. 
This state-of-the-art can also be compared with the $f(R)$ gravity, where extra scalar mode propagates. 
The corresponding scalar mode in the $f(Q)$ gravity does not propagate due to the constraint \cite{Hu:2023gui}. 
This result is consistent with that in \cite{Capozziello:2024vix}. 

In \cite{Nojiri:2024zab}, it has been proposed to define $f(Q)$ gravity theory by using the metric and four scalar fields as independent fields. 
Although the number of on-shell dynamical degrees of freedom is still not clear, the equations given by the variation of the action 
with respect to these fields become consistent. 
Note that all the equations given by the variation of the action with respect to the connection are not valid and they often conflict with each other. 
This is because the connections are constrained by the curvature-free and torsion-free conditions. 
This problem is avoided by the formulation presented in \cite{Nojiri:2024zab}. 


In this paper, we show that the symmetric teleparallel theory does not exclude the Riemann curvature constructed from the standard Levi-Civita connection. 
This idea is not new, of course. 
A model called $f(Q,B)$ gravity in \cite{Capozziello:2023vne} or $f(Q,C)$ gravity in \cite{Gadbail:2023mvu} is proposed, where $B$ or $C$ is the difference 
between $Q$ and the scalar curvature $\tilde R$ in Einstein's gravity, $B=Q-\tilde R$ or $C=\tilde R - Q$. 
For the application of $f(Q,B)$ gravity to cosmology, see \cite{Paliathanasis:2023kqs, Paliathanasis:2024yea}. 
We should note that $B$ or $C$ is a total derivative. 
The action of $f(Q, B)$ or $f(Q, C)$ gravity is given by a function $f(Q, B)$ or $f(Q, C)$ of $Q$ and $B$ or $C$. 
Even if we consider the $f(Q, \tilde R)$ gravity, whose action is given by a function $f(Q, \tilde R)$, because $f(Q, \tilde R)=f(Q, Q-B)=f(Q, Q+C)$, the model 
can be regarded as the $f(Q, B)$ or $f(Q, C)$ gravity. 
One reason why the symmetric teleparallel theory can include the curvature in Einstein's gravity is that the conservation law is usually 
given in terms of the covariant derivative defined by the Levi-Civita connection. 
The commutator of two covariant derivatives inevitably induces the curvature in Einstein's gravity. 
In this paper, we propose to consider $f \left(Q, \mathcal{G}\right) $ gravity, where $\mathcal{G}$ is the Gauss-Bonnet invariant. 
We apply this model to the cosmology and give an explicit formulation for the cosmological reconstruction. 
It is proposed a systematic way to construct such a theory which realises any given geometry. 

The paper is composed as follows: 
After the brief review of $f(Q)$ gravity in the next section~\ref{SecII}, Section~\ref{SecIII} is devoted to the explanation of why one can/should include the curvatures 
in Einstein's gravity into $f(Q)$ gravity. 
After that, we concentrate on $f \left(Q, \mathcal{G}\right) $ gravity and derive the field equations. 
After mentioning the ghost problem in the theory in Section~\ref{SecIV}, we present the explicit forms of the field equations in the 
spatially flat Friedmann-Lema\^{i}tre-Robertson-Walker (FLRW) spacetime in Section~\ref{SecV}. 
In this section, we also investigate the formalism of the cosmological reconstruction, which realises an arbitrary given FLRW spacetime. 
In this formalism, we are interested in finding a model that realises the geometry desired from the theoretical and/or observational viewpoints. 
If the model can consistently realise the spacetime, the model can be realistic gravity theory. 
In Section~\ref{SecVI}, we apply the formalism of the cosmological reconstruction and construct models which mimic the $\Lambda$CDM model, 
describe the slow-roll or constant-roll 
inflation, unify inflation and dark energy. Finally, we consider a dynamical autonomous system in the model. 
Some remarks on the gravitational wave in $f\left(Q,\mathcal{G}\right)$ are also made. 
After that, in Section~\ref{SecVII}, we briefly note possible generalisation of the model like $f\left( Q, \tilde R, \mathcal{G} \right)$ theory. 
The last section is devoted to Summary and Discussions.

\section{Brief review of $f(Q)$ gravity\label{SecII}}

As is mentioned in \cite{Nojiri:2024zab}, the treatment of the $f(Q)$ gravity is often not well-defined. 
This comes from the problem of the dynamical degrees of freedom. 
One cannot use all the equations given by the variation of the action with respect to the connection because 
the connection is very restricted by requiring that the torsion and the Riemann tensor should vanish. Therefore all the components of the connection are not independent. 
In \cite{Nojiri:2024zab}, we defined the $f(Q)$ model only by using the metric and four scalar fields $\xi^a$. 
As a result, the model under consideration becomes well-defined and the inconsistency in choosing the so-called coincident gauge in the FLRW spacetime 
becomes clear. 



The general affine connection can be decomposed into three parts, 
\begin{align}
\label{affine}
{\Gamma^\sigma}_{\mu \nu}= {\tilde\Gamma^\sigma}_{\mu\nu} + {K^\sigma}_{\mu \nu} + {L^\sigma}_{\mu \nu}\,.
\end{align}
Here ${\tilde\Gamma^\sigma}_{\mu\nu}$ is the standard Levi-Civita connection in the Einstein gravity, 
\begin{align}
\label{Levi-Civita}
{{\tilde\Gamma}^\sigma}_{\mu \nu} = \frac{1}{2} g^{\sigma \rho} \left( \partial_\mu g_{\rho \nu} + \partial_\nu g_{\rho \mu}- \partial_\rho g_{\mu \nu}\right)\, ,
\end{align}
${K^\sigma}_{\mu \nu}$ is called contortion and given by the anti-symmetric part of the connection, that is, torsion tensor, 
${T^\sigma}_{\mu \nu}={\Gamma^\sigma}_{\mu \nu} - {\Gamma^\sigma}_{\nu \mu}$, 
\begin{align}
\label{contortion}
{K^\sigma}_{\mu \nu}= \frac{1}{2} \left( {T^\sigma}_{\mu \nu} + T^{\ \sigma}_{\mu\ \nu} + T^{\ \sigma}_{\nu\ \mu} \right) \, ,
\end{align}
and ${L^\sigma}_{\mu \nu}$ is named as deformation and given by, 
\begin{align}
\label{deformation}
{L^\sigma}_{\mu \nu}= \frac{1}{2} \left( Q^\sigma_{\;\mu \nu} - Q^{\ \sigma}_{\mu\ \nu} - Q^{\ \sigma}_{\nu\ \mu} \right)\,.
\end{align}
We call ${Q^\sigma}_{\mu \nu}$ non-metricity tensor and define it by, 
\begin{align}
\label{non-metricity}
Q_{\sigma \mu \nu}= \nabla_\sigma g_{\mu \nu}= \partial_\sigma g_{\mu \nu} - {\Gamma^\rho}_{\sigma \mu } g_{\nu \rho} - {\Gamma^\rho}_{\sigma \nu } g_{\mu \rho } \,.
\end{align}
The tensor $Q_{\sigma \mu \nu}$ is used for the construction of the $f(Q)$ gravity. 

We only consider the case without torsion and assume ${\Gamma^\sigma}_{\mu\nu}$ is symmetric under the exchange of $\mu$ and $\nu$, 
${\Gamma^\sigma}_{\mu\nu} = {\Gamma^\sigma}_{\nu\mu}$. 
By requiring the Riemann tensor to vanish, 
\begin{align}
\label{curvatures}
R^\lambda_{\ \mu\rho\nu} \equiv {\Gamma^\lambda}_{\mu\nu,\rho} - {\Gamma^\lambda}_{\mu\rho,\nu} 
+ {\Gamma^\eta}_{\mu\nu}{\Gamma^\lambda}_{\rho\eta} - {\Gamma^\eta}_{\mu\rho}{\Gamma^\lambda}_{\nu\eta} =0 \, ,
\end{align}
we obtain symmetric teleparallel theories of gravity. 
A solution of Eq.~(\ref{curvatures}) is given by using four fields $\xi^a$ $\left( a = 0,1,2,3 \right)$ as follows, 
\begin{align}
\label{G1B}
{\Gamma^\rho}_{\mu\nu}=\frac{\partial x^\rho}{\partial \xi^a} \partial_\mu \partial_\nu \xi^a \, .
\end{align}
In the Levi-Civita connection (\ref{Levi-Civita}), this connection (\ref{G1B}) corresponds to the pure gauge, 
that is, the connection is given by the coordinate transformation $x^a\to \xi^a$ from the flat spacetime, whose metric is given by 
$\eta_{ab}={\tiny \left( \begin{array}{cccc} -1 & 0 & 0 & 0 \\ 0 & 1 & 0 & 0 \\ 0 & 0 & 1 & 0 \\ 0 & 0 & 0& 1 \end{array} \right)}$, 
$g_{\mu\nu}= \frac{\partial \xi^a}{\partial x^\mu} \frac{\partial \xi^b}{\partial x^\nu} \eta_{ab}$. 
As the curvature is covariant under the coordinate transformation, the obtained curvature vanishes because the curvature vanishes in the flat spacetime. 

As we see later, $\xi^a$'s should not be vector fields but scalar fields. 
By using gauge degrees of freedom coming from the general covariance, one may choose the gauge condition ${\Gamma^\rho}_{\mu\nu}=0$, which is called 
the coincident gauge and realised by choosing $\xi^a=x^a$. 
The gauge condition contradicts the FLRW universe and the spherically symmetric spacetime, in general, because the assumption of the FLRW universe 
or the spherically symmetric spacetime is another coordinate choice. 

Under the infinitesimal transformation of $\xi^a$, 
\begin{align}
\label{G2}
\xi^a \to \xi^a + \delta \xi^a \, .
\end{align}
the connection in (\ref{G1B}) transforms as follows, 
\begin{align}
\label{G1}
\Gamma^\rho_{\mu\nu} \to \Gamma^\rho_{\mu\nu} + \delta \Gamma^\rho_{\mu\nu}
\equiv \Gamma^\rho_{\mu\nu} - \frac{\partial x^\rho}{\partial \xi^a} \partial_\sigma \delta\xi^a\frac{\partial x^\sigma}{\partial \xi^b}\partial_\mu \partial_\nu \xi^b
+ \frac{\partial x^\rho}{\partial \xi^a} \partial_\mu \partial_\nu \delta \xi^a \, .
\end{align}
If we regard $\xi^a$'s as scalar fields, 
we obtain $\delta\xi^a = \epsilon^\mu \partial_\mu \xi^a$ under the coordinate transformation $x^\mu\to x^\mu + \epsilon^\mu$ with infinitesimally small functions $\epsilon^\mu$. 
This shows, 
\begin{align}
\label{G1GCT}
\delta \Gamma^\rho_{\mu\nu}
= \epsilon^\sigma \partial_\sigma \Gamma^\rho_{\mu\nu} - \partial_\sigma \epsilon^\rho \Gamma^\sigma_{\mu\nu} 
+ \partial_\mu \epsilon^\eta \Gamma^\rho_{\eta\nu} + \partial_\nu \epsilon^\eta \Gamma^\rho_{\mu\eta} 
+ \partial_\mu \partial_\nu \epsilon^\rho \, .
\end{align}
The last term in (\ref{G1GCT}) is the inhomogeneous term, by which the general covariance of the covariant derivative is guaranteed. 
This is approved because $\xi^a$'s are scalar fields. 

The symmetric teleparallel theory is given by the scalar of the non-metricity, which is defined as follows, 
\begin{align}
\label{non-m scalar}
Q\equiv g^{\mu \nu} \left( {L^\alpha}_{\beta \nu}{L^\beta}_{\mu \alpha} - {L^\beta}_{\alpha \beta} {L^\alpha}_{\mu \nu} \right)
 -Q_{\sigma \mu \nu} P^{\sigma \mu \nu} \, ,
\end{align}
with the following definitions of ${L^\sigma}_{\mu \nu}$ and $P^{\sigma \mu \nu}$, 
\begin{align}
\label{deformationB}
{L^\sigma}_{\mu \nu}\equiv &\, \frac{1}{2} \left( Q^\sigma_{\;\mu \nu} - Q^{\ \sigma}_{\mu\ \nu} - Q^{\ \sigma}_{\nu\ \mu} \right)\, , \\
\label{non-m conjugate}
{P^\sigma}_{\mu \nu} \equiv &\, \frac{1}{4} \left\{ - {Q^\sigma}_{\mu \nu} + Q^{\ \sigma}_{\mu\ \nu} + Q^{\ \sigma}_{\nu\ \mu}
+ Q^\sigma g_{\mu \nu}- \tilde{Q}^\sigma g_{\mu \nu} - \frac{1}{2} \left( {\delta^\sigma}_\mu Q_\nu + {\delta^\sigma}_\nu Q_\mu \right) \right\}\, , \nonumber \\
Q_\sigma \equiv&\, Q^{\ \mu}_{\sigma\ \mu}\, , \quad \tilde{Q}_\sigma=Q^\mu_{\ \sigma \mu}\, ,
\end{align}
Then 
\begin{align}
\label{Q}
Q=&\, - \frac{1}{4} g^{\alpha\mu} g^{\beta\nu} g^{\gamma\rho} \nabla_\alpha g_{\beta\gamma} \nabla_\mu g_{\nu\rho}
+ \frac{1}{2} g^{\alpha\mu} g^{\beta\nu} g^{\gamma\rho} \nabla_\alpha g_{\beta\gamma} \nabla_\rho g_{\nu\mu}
+ \frac{1}{4} g^{\alpha\mu} g^{\beta\gamma} g^{\nu\rho} \nabla_\alpha g_{\beta\gamma} \nabla_\mu g_{\nu\rho} \nonumber \\
&\, - \frac{1}{2} g^{\alpha\mu} g^{\beta\gamma} g^{\nu\rho} \nabla_\alpha g_{\beta\gamma} \nabla_\nu g_{\mu\rho} \, .
\end{align}

The $f(Q)$ gravity is given by the following action, 
\begin{align}
\label{ActionQ}
S=\int d^4 x \sqrt{-g} f(Q)\, .
\end{align}
There is a problem with the number of degrees of freedom in $f(Q)$ gravity. 
Although the functional degrees of freedom in the connection are restricted by the curvature-free and torsion-free conditions, 
it is not so clear which could be valid in the equations given by the variation with respect to the connection. 
In order to avoid this problem, we regard the metric $g_{\mu\nu}$ and $\xi^a$ as independent fields as in \cite{Nojiri:2024zab}. 

\section{$Q\mathcal{G}$ gravity\label{SecIII}}

In this section, we show that non-metricity gravity does not exclude the Riemann curvatures given by the Levi-Civita connection in (\ref{Levi-Civita}). 
First, we show that the covariant derivative defined in the Levi-Civita connection (\ref{Levi-Civita}) must appear even in the $f(Q)$ gravity when we include matter. 

One should remember that the conservation of matter is obtained only by using the matter equation of motion. 
Even if we consider other kind of gravity theories, we find the conservation law of the matter energy-momentum tensor $T_{\mu\nu}$. 
Therefore the conservation should be described by the Levi-Civita connection ${\tilde \Gamma^\sigma}_{\mu \nu}$ in (\ref{Levi-Civita}) for Einstein's gravity 
even if one considers the $f(Q)$ gravity theory, 
\begin{align}
\label{cons2}
0 = \tilde\nabla^\mu T_{\mu\nu} = g^{\rho\mu} \left( \partial_\rho T_{\mu\nu} - \tilde \Gamma^\sigma_{\; \rho\mu} T_{\sigma\nu} 
 - \tilde \Gamma^\sigma_{\; \rho\nu} T_{\sigma\mu} \right) \, .
\end{align}
Because the equation corresponding to the Einstein equation is given by 
\begin{align}
\label{EQeq}
0= \mathcal{G}_{\mu\nu} + T_{\mu\nu}\, , \quad \mathcal{G}_{\mu\nu} \equiv \frac{1}{\sqrt{-g}} g_{\mu\rho} g_{\nu\sigma}\frac{\delta S}{\delta g_{\rho\sigma}} \, ,
\end{align}
by the consistency of the Lagrangian theory, which is a condition of functional integrability, one gets, 
\begin{align}
\label{QBianchi}
0 = \tilde\nabla^\mu \mathcal{G}_{\mu\nu} \, ,
\end{align}
This is the generalised Bianchi identity in $f(Q)$ gravity theory. 

Therefore the covariant derivative given by only the Levi-Civita connection ${\tilde \Gamma^\sigma}_{\mu \nu}$ appears even in the $f(Q)$ gravity. 
The commutator of the Levi-Civita covariant derivatives gives the Riemann curvature in the Einstein gravity. 
The curvatures do not, of course, conflict with the general covariance. 
Therefore one may consider the model of $f(Q)$ gravity including the curvatures in the Einstein gravity, where the Lagrangian density 
is given by the function of $Q$ and the scalar curvature in the Einstein gravity $\tilde R$, that is, $f(Q, \tilde R)$, or even 
the Gauss-Bonnet invariant $\mathcal{G}$, $f(Q, \tilde R, \mathcal{G})$, whose action is given by 
\begin{align}
\label{Gnrl1QRG}
S_{\mathrm{QR}\mathcal{G}} = \int d^4x \sqrt{-g} f\left(Q, \tilde R, \mathcal{G} \right)\, .
\end{align}
where the Gauss-Bonnet invariant $\mathcal{G}$ is defined by 
\begin{align}
\label{GB} 
\mathcal{G}={\tilde R}^2 -4 {\tilde R}_{\mu\nu}
{\tilde R}^{\mu\nu} + {\tilde R}_{\mu\nu\xi\sigma} {\tilde R}^{\mu\nu\xi\sigma}\, . 
\end{align}
The $f(Q, \tilde R)$ gravity is equivalent to $f(Q,C)$ gravity proposed in \cite{Gadbail:2023mvu} because $C$ is the difference between $Q$ and the scalar curvature $\tilde R$, 
$C=\tilde R - Q$. 

Because $f(R)$ gravity includes the scalar mode, $f(Q, \tilde R, \mathcal{G})$ could also include the scalar mode. 
For this reason, in the following, we consider a simpler model, that is, $Q\mathcal{G}$ gravity, 
where the Lagrangian density is a function of $Q$ and $\mathcal{G}$, $f(Q, \mathcal{G})$, 
\begin{align}
\label{fQG1}
S_{Q\mathcal{G}}= \int d^4 x \sqrt{-g} f\left(Q, \mathcal{G} \right) \, .
\end{align}
Then by the variation of the action with respect to the metric, one obtains
\begin{align}
\label{fQG2}
0=&\, T_{\mu\nu} + \frac{1}{2} g_{\mu\nu} f 
 - f_Q g^{\alpha\beta} g^{\gamma\rho} \left\{ -\frac{1}{4} \nabla_\mu g_{\alpha\gamma} \nabla_\nu g_{\beta\rho}
 - \frac{1}{2} \nabla_\alpha g_{\mu\gamma} \nabla_\beta g_{\nu\rho} \right. \nonumber \\
&\, + \frac{1}{2} \left( \nabla_\mu g_{\alpha\gamma} \nabla_\rho g_{\beta\nu}
+ \nabla_\nu g_{\alpha\gamma} \nabla_\rho g_{\beta\mu} \right)
+ \frac{1}{2} \nabla_\alpha g_{\mu\gamma} \nabla_\rho g_{\nu\beta}
+ \frac{1}{4} \nabla_\mu g_{\alpha\beta} \nabla_\nu g_{\gamma\rho}
+ \frac{1}{2} \nabla_\alpha g_{\mu\nu} \nabla_\beta g_{\gamma\rho} \nonumber \\
&\, \left. - \frac{1}{4} \left( \nabla_\mu g_{\alpha\beta} \nabla_\gamma g_{\nu\rho} + \nabla_\nu g_{\alpha\beta} \nabla_\gamma g_{\mu\rho} \right)
 - \frac{1}{2} \nabla_\alpha g_{\mu\nu} \nabla_\gamma g_{\beta\rho}
 - \frac{1} {4} \left(\nabla_\alpha g_{\gamma\rho} \nabla_\mu g_{\beta\nu} + \nabla_\alpha g_{\gamma\rho} \nabla_\nu g_{\beta\mu} \right)
\right\} \nonumber \\
&\, - \frac{g_{\mu\rho} g_{\nu\sigma}}{\sqrt{-g}} \partial_\alpha \left[ \sqrt{-g} f_Q \left\{ - \frac{1}{2} g^{\alpha\beta} g^{\gamma\rho} g^{\sigma\tau} \nabla_\beta g_{\gamma\tau}
+ \frac{1}{2} g^{\alpha\beta} g^{\gamma\rho} g^{\sigma\tau} \left( \nabla_\tau g_{\gamma\beta} + \nabla_\gamma g_{\tau\beta} \right) \right. \right. \nonumber \\
&\, \left. \left. + \frac{1}{2} g^{\alpha\beta} g^{\rho\sigma} g^{\gamma\tau} \nabla_\beta g_{\gamma\tau}
 - \frac{1}{2} g^{\alpha\beta} g^{\rho\sigma} g^{\gamma\tau} \nabla_\gamma g_{\beta\tau}
 - \frac{1}{4} \left( g^{\alpha\sigma} g^{\gamma\tau} g^{\beta\rho} + g^{\alpha\rho} g^{\gamma\tau} g^{\beta\sigma}
\right) \nabla_\beta g_{\gamma\tau}
\right\} \right] \nonumber \\
&\, - \frac{g_{\mu\rho} g_{\nu\sigma}}{\sqrt{-g}} {\Gamma^\rho}_{\alpha\eta} \left[ \sqrt{-g} f_Q \left\{ - \frac{1}{2} g^{\alpha\beta} g^{\gamma\eta} g^{\sigma\tau} \nabla_\beta g_{\gamma\tau}
+ \frac{1}{2} g^{\alpha\beta} g^{\gamma\eta} g^{\sigma\tau} \left( \nabla_\tau g_{\gamma\beta} + \nabla_\gamma g_{\tau\beta} \right) \right. \right. \nonumber \\
&\, \left. \left. + \frac{1}{2} g^{\alpha\beta} g^{\eta\sigma} g^{\gamma\tau} \nabla_\beta g_{\gamma\tau}
 - \frac{1}{2} g^{\alpha\beta} g^{\eta\sigma} g^{\gamma\tau} \nabla_\gamma g_{\beta\tau}
 - \frac{1}{4} \left( g^{\alpha\sigma} g^{\gamma\tau} g^{\beta\eta} + g^{\alpha\eta} g^{\gamma\tau} g^{\beta\sigma}
\right) \nabla_\beta g_{\gamma\tau}
\right\} \right] \nonumber \\
&\, - \frac{g_{\mu\rho} g_{\nu\sigma}}{\sqrt{-g}} {\Gamma^\sigma}_{\alpha\eta} \left[ \sqrt{-g} f_Q \left\{ - \frac{1}{2} g^{\alpha\beta} g^{\gamma\rho} g^{\eta\tau} \nabla_\beta g_{\gamma\tau}
+ \frac{1}{2} g^{\alpha\beta} g^{\gamma\rho} g^{\eta\tau} \left( \nabla_\tau g_{\gamma\beta} + \nabla_\gamma g_{\tau\beta} \right) \right. \right. \nonumber \\
&\, \left. \left. + \frac{1}{2} g^{\alpha\beta} g^{\rho\eta} g^{\gamma\tau} \nabla_\beta g_{\gamma\tau}
 - \frac{1}{2} g^{\alpha\beta} g^{\rho\eta} g^{\gamma\tau} \nabla_\gamma g_{\beta\tau}
 - \frac{1}{4} \left( g^{\alpha\eta} g^{\gamma\tau} g^{\beta\rho} + g^{\alpha\rho} g^{\gamma\tau} g^{\beta\eta}
\right) \nabla_\beta g_{\gamma\tau}
\right\} \right] \nonumber \\
&\, -2 f_\mathcal{G} {\tilde R} {\tilde R}_{\mu\nu} 
+ 4f_\mathcal{G} {\tilde R}_{\mu\rho} {\tilde R}_\nu^{\ \rho} 
 -2 f_\mathcal{G} {\tilde R}_\mu^{\ \rho\sigma\tau} {\tilde R}_{\nu\rho\sigma\tau} \nonumber \\
&\, - 4 f_\mathcal{G} {\tilde R}_{\mu\rho\sigma\nu} {\tilde R}^{\rho\sigma} 
+ 2 \left( \tilde\nabla_\mu \tilde\nabla_\nu f_\mathcal{G} \right){\tilde R} \nonumber \\ 
&\, - 2 g_{\mu\nu} \left( \tilde\nabla^2 f_\mathcal{G} \right) {\tilde R} 
 - 4 \left( \tilde\nabla^\rho \tilde\nabla_\mu f_\mathcal{G} \right) {\tilde R}_{\nu\rho} 
 - 4 \left( \tilde\nabla^\rho \tilde\nabla_\nu f_\mathcal{G} \right){\tilde R}_{\mu\rho} \nonumber \\
&\, + 4 \left( \tilde\nabla^2 f_\mathcal{G} \right){\tilde R}_{\mu\nu} 
+ 4g_{\mu\nu} \left( \tilde\nabla_{\rho} \tilde\nabla_\sigma f_\mathcal{G} \right) {\tilde R}^{\rho\sigma}
 - 4 \left(\tilde\nabla^\rho \tilde\nabla^\sigma f_\mathcal{G} \right) {\tilde R}_{\mu\rho\nu\sigma} \, .
\end{align}
Here $f_Q \equiv \frac{\partial f }{\partial Q}$ and 
$f_\mathcal{G} \equiv \frac{\partial f }{\partial \mathcal{G}}$. 

We assume the non-metricity connection in (\ref{G1B}) or four scalar fields $\xi^a$ do not couple directly with the matter. 
Then the variation of the action with respect to $\xi^a$ is given by 
\begin{align}
\label{eq3}
0=&\, X_a \equiv \frac{1}{\sqrt{-g}} \frac{\delta S}{\delta \xi^a} 
= \partial_\sigma
\left\{ \frac{\partial x^\eta}{\partial \xi^a} \frac{\partial x^\sigma}{\partial \xi^b}\partial_\xi \partial_\zeta \xi^b 
\left( \sqrt{-g} \mathcal{H}^{\xi\zeta}_\eta \right) \right\}
+ \partial_\xi \partial_\zeta \left\{ \frac{\partial x^\eta}{\partial \xi^a} \left( \sqrt{-g} \mathcal{H}^{\xi\zeta}_\eta \right) \right\} 
\nonumber \\
=&\, - \partial_\sigma
\left[ \frac{\partial x^\eta}{\partial \xi^a} \frac{\partial x^\sigma}{\partial \xi^b}\partial_\xi \partial_\zeta \xi^b \left\{ \sqrt{-g} f_Q 
 \left( 
 - \frac{1}{2} g^{\xi\rho} g^{\zeta\nu} g^{\gamma\mu} 
 - \frac{1}{2} g^{\xi\rho} g^{\gamma\nu} g^{\zeta\mu} 
+ 4 g^{\xi\mu} g^{\zeta\nu} g^{\gamma\rho} \right. \right. \right. \nonumber \\
&\, \qquad \qquad \quad \left. \left. \left. + \frac{1}{4} g^{\xi\mu} g^{\zeta\gamma} g^{\nu\rho} 
 - g^{\xi\nu} g^{\zeta\gamma} g^{\mu\rho} 
 - g^{\mu\zeta} g^{\nu\rho} g^{\xi\gamma} 
 - g^{\mu\gamma} g^{\nu\rho} g^{\xi\zeta} 
\right) 
g_{\eta\gamma} \nabla_\mu g_{\nu\rho} \right\} \right] \nonumber \\
&\, - \partial_\xi \partial_\zeta \left[\frac{\partial x^\eta}{\partial \xi^a} \left\{ \sqrt{-g} f_Q 
 \left( 
 - \frac{1}{2} g^{\xi\rho} g^{\zeta\nu} g^{\gamma\mu} 
 - \frac{1}{2} g^{\xi\rho} g^{\gamma\nu} g^{\zeta\mu} 
+ 4 g^{\xi\mu} g^{\zeta\nu} g^{\gamma\rho} \right. \right. \right. \nonumber \\
&\, \qquad \qquad \quad \left. \left. \left. + \frac{1}{4} g^{\xi\mu} g^{\zeta\gamma} g^{\nu\rho} 
 - g^{\xi\nu} g^{\zeta\gamma} g^{\mu\rho} 
 - g^{\mu\zeta} g^{\nu\rho} g^{\xi\gamma} 
 - g^{\mu\gamma} g^{\nu\rho} g^{\xi\zeta} 
\right) 
g_{\eta\gamma} \nabla_\mu g_{\nu\rho} \right\} \right] \, .
\end{align}
Here
\begin{align}
\label{eq2}
\mathcal{H}^{\xi\zeta}_\eta \equiv&\, \frac{1}{\sqrt{-g}} \frac{\delta S}{\delta {\Gamma^\eta}_{\xi\zeta}} \nonumber \\
=&\, - f_Q \left( 
 - \frac{1}{2} g^{\xi\rho} g^{\zeta\nu} g^{\gamma\mu} 
 - \frac{1}{2} g^{\xi\rho} g^{\gamma\nu} g^{\zeta\mu} 
+ 2 g^{\xi\mu} g^{\zeta\nu} g^{\gamma\rho} \right. \nonumber \\
&\, \qquad \qquad \quad \left. + \frac{1}{4} g^{\xi\mu} g^{\zeta\gamma} g^{\nu\rho} 
 - g^{\xi\nu} g^{\zeta\gamma} g^{\mu\rho} 
 - \frac{1}{2} g^{\mu\zeta} g^{\nu\rho} g^{\xi\gamma} 
 - \frac{1}{2} g^{\mu\gamma} g^{\nu\rho} g^{\xi\zeta} 
\right) g_{\eta\gamma} \nabla_\mu g_{\nu\rho} \, .
\end{align}
Therefore we obtained the closed set of the equations which we should solve. 
Note that one cannot put $\mathcal{H}^{\xi\zeta}_\eta=0$ because all the components of the connections are not valid 
due to the constraints of the absence of curvatures (\ref{curvatures}) and the torsionless $T^\sigma_{\ \mu\nu}=0$. 

\section{Ghost problem\label{SecIV}}

Let us remark that the Einstein-$f\left( \mathcal{G} \right)$ gravity includes ghosts in general. Nevertheless, some ghost-free models were proposed in \cite{Nojiri:2018ouv}. 
In $f(Q, \tilde R)$ or $\mathcal{G}$, $f(Q, \tilde R, \mathcal{G})$, there could appear the ghost. 
As well known, the kinetic energy of the ghosts is unbounded below and there occur instabilities of the system in the classical theory.
In the quantum theory, as in the Fadeev-Popov ghosts in the gauge theories~\cite{Kugo:1979gm}, the ghosts generate the negative norm states, 
which tells that negative probabilities appear and therefore the ghosts violate the Copenhagen interpretation of the quantum theory.
The existence of the ghost tells that the model is physically inconsistent and therefore the problem of the ghosts is very important. 
If we find any ghosts in any model, we need to abandon the model or modify the model not to include ghosts.
First, we briefly review the ghost problem in the Einstein-$f\left( \mathcal{G} \right)$ gravity.

The action of the Einstein-$f\left( \mathcal{G} \right)$ gravity~\cite{Nojiri:2005jg, Cognola:2006eg, Leith:2007bu, Li:2007jm, Kofinas:2014owa, Zhou:2009cy, Oikonomou:2015qha} 
 is given by 
\begin{align}
\label{GB1b} 
S=\int d^4x\sqrt{-g} \left(\frac{1}{2\kappa^2}{\tilde R} +f(\mathcal{G}) + \mathcal{L}_\mathrm{matter}\right)\, .
\end{align}
This model is motivated by the scalar-Einstein-Gauss-Bonnet gravity theory~\cite{Nojiri:2005vv,Nojiri:2006je}, whose action is given by, 
\begin{align}
\label{FRGBg16}
S=\int d^4x\sqrt{-g} \left(\frac{1}{2\kappa^2}{\tilde R}
 - \frac{1}{2} \partial_\mu \chi \partial^\mu \chi
+ h\left( \chi \right) \mathcal{G} - V\left( \chi \right) + \mathcal{L}_\mathrm{matter}\right)\, . 
\end{align}
Here $\chi$ is a scalar field, $h(\chi)$ is a scalar potential coupling with the Gauss-Bonnet invariant $\mathcal{G}$, and $V(\chi)$ is the potential of the scalar field $\chi$. 
The action of the Einstein-$f\left( \mathcal{G} \right)$ gravity (\ref{GB1b}) can be obtained by dropping the kinetic term of the scalar field $\chi$, 
\begin{align}
\label{FRGBg12}
S=\int d^4x\sqrt{-g} \left(\frac{1}{2\kappa^2}{\tilde R}
+ h\left( \chi \right) \mathcal{G} - V\left( \chi \right)
+ \mathcal{L}_\mathrm{matter}\right)\, .
\end{align}
By the variation of the action (\ref{FRGBg12}) with respect to the auxiliary field $\chi$, 
we obtain the following equation,
\begin{align}
\label{FRGBg13}
0 = h'\left( \chi \right) \mathcal{G} - V'\left( \chi \right) \, ,
\end{align}
which can be solved with respect to $\chi$ as a function of the Gauss-Bonnet invariant $\mathcal{G}$, $\chi = \chi\left( \mathcal{G} \right)$. 
By substituting the obtained expression into Eq.~(\ref{FRGBg13}), one reobtains the action of Eq.~(\ref{GB1b}) with $f \left( \mathcal{G} \right)$ being equal to,
\begin{align}
\label{FRGBg14}
f \left( \mathcal{G} \right) = h \left( \chi \left( \mathcal{G} \right) \right) \mathcal{G}
 - V \left( \chi \left( \mathcal{G} \right) \right) \, .
\end{align}
This may also indicate that Einstein-$f\left( \mathcal{G} \right)$ gravity could be also obtained in some limit of 
the scalar-Einstein-Gauss-Bonnet gravity which is often regarded as string-inspired gravity.

In \cite{Nojiri:2018ouv}, it has been shown that under the perturbation $g_{\mu\nu}=g^{(0)}_{\mu\nu} + \delta g_{\mu\nu}$, 
the field equation includes the fourth derivative of the metric $g_{\mu\nu}$ with respect to the time coordinate, 
and therefore the perturbed equation generates a ghost mode, which is a scalar. 
On the other hand, we can easily find that the scalar-Einstein-Gauss-Bonnet gravity (\ref{FRGBg16}) does not include ghosts.

In the model of Eq.~(\ref{FRGBg16}), there appears one more dynamical degree of freedom, namely $\chi$, compared with 
the Einstein-$f\left( \mathcal{G} \right)$ gravity (\ref{GB1b}). 
In order to reduce the dynamical degrees of freedom, a constraint as in the mimetic gravity 
case~\cite{Chamseddine:2013kea, Nojiri:2014zqa,Dutta:2017fjw} is imposed in \cite{Nojiri:2018ouv}, by
using the Lagrange multiplier field $\lambda$, 
\begin{align}
\label{FRGBg19}
S=\int d^4x\sqrt{-g} \left\{ \frac{1}{2\kappa^2}{\tilde R}
+ \lambda \left( \frac{1}{2} \partial_\mu \chi \partial^\mu \chi + \frac{\mu^4}{2} \right)
 - \frac{1}{2} \partial_\mu \chi \partial^\mu \chi
+ h\left( \chi \right) \mathcal{G} - V\left( \chi \right) + \mathcal{L}_\mathrm{matter}\right\}\, . 
\end{align}
Here $\mu$ is a constant with mass dimension. 
By the variation of the above action (\ref{FRGBg19}) with respect to $\lambda$, the following constraint is obtained,
\begin{align}
\label{FRGBg20}
0=\frac{1}{2} \partial_\mu \chi \partial^\mu \chi + \frac{\mu^4}{2} \, .
\end{align}
It has been shown that due to this constraint, the scalar field $\chi$ becomes non-dynamical, that is, $\chi$ does not propagate, 
in addition to the absence of ghosts. 
Even in the case of $Q\mathcal{G}$ gravity (\ref{fQG1}), the ghosts coming from $\mathcal{G}$ sector could be eliminated by using the formulation similar to the case of 
the Einstein-$f\left( \mathcal{G} \right)$ gravity. 
In fact, even for $f\left( Q, \mathcal{G} \right)$ gravity, it is not difficult to consider analogous models including a scalar field, 
\begin{align}
\label{FRGBg12Q}
S=\int d^4x\sqrt{-g} \left( f(Q)
+ h\left( \chi \right) \mathcal{G} - V\left( \chi \right)
+ \mathcal{L}_\mathrm{matter}\right)\, ,
\end{align}
which is a scalar--$f(Q)$--Gauss-Bonnet gravity. As in (\ref{FRGBg19}), we may introduce the Lagrange multiplier field, 
\begin{align}
\label{FRGBg19Q}
S=\int d^4x\sqrt{-g} \left\{ f(Q) + \lambda \left( \frac{1}{2} \partial_\mu \chi \partial^\mu \chi + \frac{\mu^4}{2} \right)
 - \frac{1}{2} \partial_\mu \chi \partial^\mu \chi
+ h\left( \chi \right) \mathcal{G} - V\left( \chi \right) + \mathcal{L}_\mathrm{matter}\right\}\, . 
\end{align}
In these models (\ref{FRGBg12Q}) and (\ref{FRGBg19Q}), the ghosts related to the Gauss-Bonnet invariant $\mathcal{G}$ do not appear. 

At present, it is difficult to discuss the ghost problem in $f(Q, \tilde R)$ or $\mathcal{G}$, $f(Q, \tilde R, \mathcal{G})$ gravity theory. 
This is because the problem of the dynamical degrees of freedom (DOF) of $f(Q)$ gravity has not been solved \cite{Hu:2022anq, DAmbrosio:2023asf, Heisenberg:2023lru, 
Paliathanasis:2023pqp, Dimakis:2021gby, Hu:2023gui}. 
Hence, we cannot exclude the possibility that there might appear ghost in the $Q$-sector. 
The study of the propagating mode in the flat background could tell that the only DOF could be a graviton as in Einstein's theory~\cite{Capozziello:2024vix}, where the 
situation is similar to $f(T)$ gravity \cite{Bamba:2013ooa}. 
Note that in $f(R)$ gravity, there appears an extra scalar mode, which appears as a scale of the metric 
of the scalar curvature, but this scalar mode is not a ghost. 
In the case of the $f(Q)$ gravity, the corresponding scalar mode seems to behave as a ghost but it has been shown that this mode does not propagate 
due to the constraint \cite{Hu:2023gui}. 
This result is consistent with that in \cite{Capozziello:2024vix}. 
Even in the case of the $f(T)$ gravity, the only propagating mode in the flat background is a graviton \cite{Bamba:2013ooa} but 
it is known that in the $f(T)$ gravity, there appear superluminal propagating modes, which may indicate that the $f(T)$ gravity theory could be inconsistent~\cite{Ong:2013qja, Izumi:2012qj}. 
In the case of the $f(Q)$ gravity, there might be a similar mode in the higher-order perturbation. 
If one can separate such an unphysical mode, the mode could be eliminated by using the constraint as in (\ref{FRGBg20}).



\section{FLRW spacetime in $f\left(Q, \mathcal{G} \right)$ gravity and cosmological reconstruction\label{SecV}}

In this section, we work in the spatially flat FLRW spacetime whose metric is given, 
\begin{align}
\label{FLRW}
ds^2 = - dt^2 + a(t)^2 \sum_{i=1,2,3} \left( dx^i \right)^2 \, .
\end{align}
Here $t$ is the cosmological time and $a(t)$ is called a scale factor. 
We also assume
\begin{align}
\label{xi}
\xi^0 = b(t)\, , \quad \xi^i = x^i\, , 
\end{align}
which gives
\begin{align}
\label{connctn}
{\Gamma^0}_{00}=\gamma(t) \equiv \frac{\ddot b(t)}{\dot b(t)} \, , \quad \mbox{other components of the connection}=0\, . 
\end{align}
The assumption (\ref{xi}) could be an almost unique choice which does not violate the rotational symmetry in the spatial part. 
Under the assumption, Eq.~(\ref{eq2}) gives, 
\begin{align}
\label{H000}
\mathcal{H}_0^{00} =
\frac{3}{2} f_Q \left(Q, \mathcal{G} \right) \left( \gamma - H \right) \, , 
\quad 
\mathcal{H}_0^{ij} =
 - f_Q \left(Q, \mathcal{G} \right) a^{-2} \delta^{ij} \left( - 2 H + \gamma \right) \, , 
\quad
\mathcal{H}_j^{0i} =
\frac{f_Q \left(Q, \mathcal{G} \right)}{2} \delta_i^{\ j} \left( 7 H + 5 \gamma \right) \, , 
\end{align}
and therefore Eq.~(\ref{eq3}) is
\begin{align}
\label{Xa}
0= X_0 = \frac{d}{dt} \left[ \left\{\frac{3\gamma a^3}{2 b'} f_Q \left(Q, \mathcal{G} \right) \left( \gamma - H \right) \right\} \right] 
+ \frac{d^2}{dt^2} \left\{ \frac{a^3}{b'} f_Q \left(Q, \mathcal{G} \right) \left( \gamma - H \right) \right\} \, . 
\end{align}
Note that $X_i$ vanishes identically. 
The equations in (\ref{H000}) 
tell that one cannot put $\mathcal{H}^{\xi\zeta}_\eta=0$. 
As we mentioned, this is because all the components of the connections are not valid due to the constraints ${\tilde R}^\lambda_{\ \mu\rho\nu}=0$ 
(\ref{curvatures}) and the torsionless condition $T^\sigma_{\ \mu\nu}=0$. 
 From Eq.~(\ref{Xa}), we find, 
\begin{align}
\label{gammaH}
\gamma=H \, .
\end{align}
Therefore the coincident gauge, where $\gamma=0$, is not consistent with the FLRW spacetime. 

On the other hand, Eq.~(\ref{fQG2}) gives, 
\begin{align}
\label{G00}
0 =&\, - f - 12 H^2 f_Q 
+ \mathcal{G}f_\mathcal{G} - 24 \dot{\mathcal{G}}f_{\mathcal{G}\mathcal{G}} H^3 + \rho \, , \\
\label{Gij}
0=&\, f + 4 a^{-3} \frac{d}{dt} \left( a^3 H f_Q \right) 
 - \mathcal{G}f_\mathcal{G} 
+ \frac{2 \mathcal{G} \dot{\mathcal{G}}}{3H} f_{\mathcal{G}\mathcal{G}} 
+ 8 H^2 \ddot{\mathcal{G}} f_{\mathcal{G}\mathcal{G}} 
+ 8 H^2 {\dot{\mathcal{G}}}^2 f_{\mathcal{G}\mathcal{G}\mathcal{G}} + p\, .
\end{align}
Here $\rho$ and $p$ are the energy density and the pressure of matter, respectively. 
We should note that Eq.~(\ref{Gij}) can be obtained from (\ref{G00}) and the conservation law of matter, 
\begin{align}
\label{cons}
\dot\rho + 3 H \left( \rho + p \right) = 0 \, .
\end{align}
This situation results from Eq.~(\ref{QBianchi}). 
Therefore one may forget Eq.~(\ref{Gij}) and only consider Eq.~(\ref{G00}). 

We now consider the formulation of the ``reconstruction'' of the model, which realises an arbitrary given FLRW spacetime~(\ref{FLRW}). 
Let us try to find a model that realises the geometry desired from the theoretical and/or observational viewpoints. 
Assume $f\left(Q, \mathcal{G}\right)$ is given by the sum of the part including only $Q$ and that including only $\mathcal{G}$, 
\begin{align}
\label{fQG3}
f = f_1(Q) + f_2 \left( \mathcal{G} \right)\, .
\end{align}
Then Eq.~(\ref{G00}) is also separated into two parts and the part coming from matter, 
\begin{align}
\label{fQG4}
0 = - f_1\left( Q \right) - 12 H^2 f'_1\left( Q \right) 
- f_2\left( \mathcal{G} \right) + \mathcal{G}f'_2 \left( \mathcal{G} \right) - 24 \dot{\mathcal{G}}f''_2 \left( \mathcal{G} \right) H^3 + \rho \, , 
\end{align}
We should note that the FLRW spacetime~(\ref{FLRW}) includes only one functional degree of freedom, that is, $a(t)$. 
In the model (\ref{fQG3}), there appear two functional degrees of freedom $f_1(Q)$ and $f_2 \left( \mathcal{G} \right)$. 
Here we may choose $f_2 \left( \mathcal{G} \right)$ as an arbitrary function. 

Let us assume the scale factor $a$ and therefore the Hubble rate $H\equiv \frac{\dot a}{a}$ is given by a function of the cosmological time $t$, 
$a=a(t)$ and $H=H(t)$. 
Due to $Q=-6H^2$, $Q$ is also given by a function ot $t$, $Q=Q(t)$, which could be solved with respect to $t$, $t=t(Q)$. 
By using this expression, we can express $\mathcal{G}$, $\dot{\mathcal{G}}$, and $\rho$ as the functions of $Q$, $\mathcal{G}=\mathcal{G}(Q)$, 
$\dot{\mathcal{G}}=\dot{\mathcal{G}}(Q)$, and $\rho=\rho(Q)$. 
Then Eq.~(\ref{fQG4}) can be integrated to give, 
\begin{align}
\label{fQG5} 
f_1(Q) = \sqrt{-Q} \int^Q \frac{dq}{\sqrt{-q}} \left\{ - f_2\left( \mathcal{G} \left( q \right) \right) 
+ \mathcal{G} \left( q \right) f'_2 \left( \mathcal{G} \left( q \right) \right) 
+ \frac{4}{\sqrt{6}} \dot{\mathcal{G}} \left( q \right) f''_2 \left( \mathcal{G} \left( q \right) \right) \left( - q\right)^\frac{3}{2} 
+ \rho \left( q \right) \right\}\, .
\end{align}
Therefore it is obtained the theory realising the given FLRW spacetime. 


\section{Cosmological Applications in $f\left(Q, \mathcal{G}\right) $ gravity\label{SecVI}}

In this section, we apply the formalism of the reconstruction given by Eq.~(\ref{fQG5}) to the cosmology and construct models which 
mimic the $\Lambda$CDM model, realize the slow-roll or constant-roll inflation, and unify inflation and dark energy. 
Furthermore, we formulate a dynamical autonomous system in the model and we investigate the gravitational wave in $f\left(Q,\mathcal{G}\right)$. 

\subsection{Mimicking $\Lambda$CDM model}

First, we consider a model mimicking the $\Lambda$CDM model, where the scale factor is given by 
\begin{align}
\label{LCDM1}
a(t) = a_0 \sinh^\frac{2}{3} \left( \alpha t \right)\, ,
\end{align}
where $a_0$ and $\alpha$ are positive constants. 
Eq.~(\ref{LCDM1}) gives the following Hubble rate $H(t)$ and $\dot H(t)$, 
\begin{align}
\label{LCDM2}
H(t) = \frac{2}{3}\alpha \coth \left( \alpha t \right)\, . \quad 
\dot H (t) = - \frac{2}{3}\frac{\alpha^2}{\sinh^2 \left( \alpha t \right)} = \frac{2}{3} \alpha^2 - \frac{3}{2} H(t)^2 \, .
\end{align}
Therefore 
\begin{align}
\label{LCDM3}
Q(t) = - { \frac{8}{3}}\alpha^2 \coth^2 \left( \alpha t \right) \, , \quad 
a(t) = a_0 \left( Q(t) + { \frac{8}{3}}\alpha^2 \right)^{-\frac{1}{3}} \, .
\end{align}
and 
\begin{align}
\label{LCDMG}
\mathcal{G} (Q) 
= - \frac{2}{3} Q^2 - \frac{8}{3} \alpha^2 Q \, , \quad 
\dot{\mathcal{G}} (Q) 
= \frac{32}{3} \left( \alpha^2 + \frac{Q}{2} \right) \left( \alpha^2 + \frac{3}{8} Q \right) \sqrt{- \frac{Q}{6}} \, .
\end{align}
In the $\Lambda$CDM model, 
matter could be the baryonic matter with the EoS parameter $\frac{1}{3}$ and therefore 
the energy density $\rho$ is given by, 
\begin{align}
\label{LCDM5}
\rho = \rho_0 a(t)^{-3} = \frac{\rho_0}{{a_0}^3}\left( Q(t) + { \frac{8}{3}}\alpha^2 \right) \, .
\end{align}
Then Eq.~(\ref{fQG5}) shows that $f_1(Q)$ is given by 
\begin{align}
\label{fQG5LCDM} 
f_1(Q) = \sqrt{-Q} \int^Q \frac{dq}{\sqrt{-q}} \left\{ - f_2\left( \mathcal{G} \left( q \right) \right) 
+ \mathcal{G} \left( q \right) f'_2 \left( \mathcal{G} \left( q \right) \right) 
+ \frac{4}{\sqrt{6}} \dot{\mathcal{G}} \left( q \right) f''_2 \left( \mathcal{G} \left( q \right) \right) \left( - q\right)^\frac{3}{2} 
+ \frac{\rho_0}{{a_0}^3}\left( q + { \frac{8}{3}}\alpha^2 \right) \right\}\, .
\end{align}
In (\ref{fQG5LCDM}), $\mathcal{G}(q)$ and $\dot{\mathcal{G}}(q)$ are given by replacing $Q$ with $q$ in (\ref{LCDMG}). 
Then we can realise the $f\left(Q, \mathcal{G} \right)$ gravity cosmology just like in the $\Lambda$CDM model without the cosmological constant and the real dark matter. 
In (\ref{LCDM5}), we assumed the matter is only baryon but we may partially include the cold dark matter (CDM) in the matter content, which may contribute 
the structure formation. 
If there is any discrepancy between the quantity required for the expansion of the universe and that necessary for the structure formation, the model may contribute to the difference 
in an additive way or a subtractive way.

\subsection{Inflation}

Let us construct a model which realises the inflation by using Eq.~(\ref{fQG5}). 
First, we consider the slow roll inflation and we explain how one can effectively include the matter creation at the end of the inflation. 
After that, we also consider the possibility of the constant-roll inflation \cite{Inoue:2001zt, Tsamis:2003px, Kinney:2005vj, Tzirakis:2007bf, Namjoo:2012aa, 
Martin:2012pe, Motohashi:2014ppa, Cai:2016ngx, Motohashi:2017aob, Hirano:2016gmv, Anguelova:2015dgt, Cook:2015hma, Kumar:2015mfa, Odintsov:2017yud, 
Odintsov:2017qpp, Nojiri:2017qvx}.

\subsubsection{Slow roll inflation and creation of matter}

As a model of inflation, we consider the following, 
\begin{align}
\label{Hex1b}
H(t) = \frac{H_0}{1 + \alpha \ln \left( 1 + \e^{\frac{2 H_0}{\alpha} \left( t - t_0 \right)} \right)} \, .
\end{align}
This model was proposed in \cite{Nojiri:2024zab}. 
In (\ref{Hex1b}), $\alpha$ is a positive constant and $t_0$ is a constant corresponding to the time when the inflation ends. 
When $t\ll t_0$, $H$ goes to a constant $H\to H_0$, which corresponds to the inflation. 
When $t\gg t_0$, we find $H\to \frac{1}{2\left(t - t_0 \right)}$, whose behaviour expresses the radiation-dominated universe. 
In \cite{Nojiri:2024zab}, it has been shown that the constraints 
\begin{align}
\label{index}
n_s = 0.9649 \pm 0.0042\, , \quad r < 0.064\, ,
\end{align}
which were obtained by the Planck 2018 observation, can be satisfied. 

It is believed that the matter could be generated by the quantum corrections at the end of the inflation. 
These effects are not included in the classical action. 
By following \cite{Nojiri:2024zab}, we now effectively include the effects by modifying the energy density $\rho$ in (\ref{G00}) 
and the pressure $p$ in (\ref{Gij}), as follows, 
\begin{align}
\label{effrhop}
\rho \to \rho_\mathrm{eff} \equiv \rho + \mathcal{J}_\rho (Q) \, , \quad 
p \to p_\mathrm{eff} \equiv p + \mathcal{J}_p(Q)\, \, .
\end{align}
Here we assume that $\mathcal{J}_\rho(Q)$ and $\mathcal{J}_p(Q)$ are functions of $Q$ just for simplicity. 
Due to the identity (\ref{QBianchi}), $\rho_\mathrm{eff}$ and $p_\mathrm{eff}$ must satisfy the conservation law as in (\ref{cons}), 
which gives, 
\begin{align}
\label{conseff}
\dot\rho + 3 H \left( \rho + p \right) = J \equiv - \dot Q \mathcal{J}_\rho'(Q) - 3 H \left( \mathcal{J}_\rho(Q) + \mathcal{J}_p(Q) \right) \, .
\end{align}
Eq.~(\ref{conseff}) tells that $J$ can be regarded as a source of matter. 
The source $J$ is chosen not to vanish only just after the inflation and to generate matter. 

For the model (\ref{Hex1b}), by using $Q=- 6 H^2$, we obtain, 
\begin{align}
\label{Hex1b2}
\e^{\frac{2 H_0}{\alpha} \left( t - t_0 \right)} = \e^{\frac{1}{\alpha} \left(\sqrt{- \frac{{ 6}{H_0}^2}{Q}} - 1\right)} - 1 \quad \mbox{or} \quad 
t=t_0 + \frac{\alpha}{2H_0} \ln \left( \e^{\frac{1}{\alpha} \left( \sqrt{- \frac{{ 6}{H_0}^2}{Q}} - 1\right) } - 1 \right) \, .
\end{align}
Therefore $t$ is explicitly given as a function of $Q$. 

We also find 
\begin{align}
\label{derivatives}
\dot H = - \frac{2 H^2}{1 + \e^{-\frac{2 H_0}{\alpha} \left( t - t_0 \right)}} 
= 2 H^2 \left( 1 - \e^{-\frac{1}{\alpha} \left(\frac{H_0}{H} - 1 \right)} \right)
= - \frac{Q}{3} \left( 1 - \e^{-\frac{1}{\alpha} \left(\sqrt{- \frac{{ 6}{H_0}^2}{Q}}-1\right)} \right)\, . 
\end{align}
Therefore the Gauss-Bonnet invariant $\mathcal{G}$ is given by 
\begin{align}
\label{GQ1}
\mathcal{G}=24 \left( H^4 + H^2 \dot H \right) = \mathcal{G}(Q) 
\equiv 
2 Q^2 \left\{ 1 - \frac{2}{3} \e^{-\frac{1}{\alpha} \left(\sqrt{- \frac{6{H_0}^2}{Q}}-1\right)} \right\} \, .
\end{align}
We also obtain 
\begin{align}
\label{GQ2}
\dot{\mathcal{G}}
= 16 \sqrt{ - \frac{Q^5}{6}} \left[ 
1 - \frac{2}{3} \e^{-\frac{1}{\alpha} \left(\sqrt{- \frac{6{H_0}^2}{Q}}-1\right)} 
+ \frac{1}{6\alpha} \e^{-\frac{1}{\alpha} \left(\sqrt{- \frac{6{H_0}^2}{Q}}-1\right)} \sqrt{- \frac{6{H_0}^2}{Q}} \right] 
\left( 1 - \e^{-\frac{1}{\alpha} \left(\sqrt{- \frac{{ 6}{H_0}^2}{Q}}-1\right)} \right) \, .
\end{align}
 The above expressions (\ref{GQ1}) and (\ref{GQ2}) are used when we calculate the r.h.s. in (\ref{fQG5}). 

As mentioned around Eq.~(\ref{conseff}), the radiation could be generated at the end of the inflation $t\sim t_0$. 
The energy density $\rho$ of the radiation could be given as follows \cite{Nojiri:2024zab}, 
\begin{align}
\label{Hex1b3}
\rho = - \frac{Q \left( Q + 6 {H_0}^2 \right)}{2\kappa^2 \left( - Q + 6 {H_0}^2 \right)} \, .
\end{align}
When $t\ll t_0$, we find $\rho \to 0$ and when $t\gg t_0$, the behaviour of $\rho$ is given as $\rho\to - \frac{3Q}{6\kappa^2} = \frac{3H^2}{\kappa^2}$, 
which corresponds to the behaviour of Einstein's gravity. 
Because the equation of state (EoS) parameter of the radiation is $\frac{1}{3}$, the pressure $p$ is given by 
\begin{align}
\label{Hex1b3p}
p = - \frac{Q \left( Q + 6{H_0}^2 \right)}{6 \kappa^2 \left( - Q + 6{H_0}^2 \right)} \, .
\end{align}
The above $\rho$ and $p$ do not satisfy the conservation law (\ref{cons}) and we obtain 
\begin{align}
\label{consmod1}
J=&\, \dot \rho + 3 H \left( \rho + p \right) 
= \frac{H}{\kappa^2} \left\{ \frac{ 2 Q \left( - Q^2 + 12 {H_0}^2 Q + 36 {H_0}^4 \right)}{ \left( - Q + 6{H_0}^2\right)^2 } 
\left( 1 - \e^{-\frac{1}{\alpha} \left( \sqrt{- \frac{6{H_0}^2}{Q}}-1\right)} \right) 
 - \frac{Q \left( Q + 6{H_0}^2 \right)}{ - Q + 6{H_0}^2 } \right\} \, ,
\end{align}
which vanishes at the early time $t\ll t_0$ and at the late time $t\gg t_0$, as expected. 

By choosing $\mathcal{J}_\rho(Q)=0$ and using Eq.~(\ref{conseff}), it follows
\begin{align}
\label{fp}
\mathcal{J}_p = - \frac{1}{3\kappa^2} \left\{ \frac{ 2 Q \left( - Q^2 + 12 {H_0}^2 Q + 36 {H_0}^4 \right)}{ \left( - Q + 6{H_0}^2\right)^2 } 
\left( 1 - \e^{-\frac{1}{\alpha} \left( \sqrt{- \frac{6{H_0}^2}{Q}}-1\right)} \right) 
 - \frac{Q \left( Q + 6{H_0}^2 \right)}{ - Q + 6{H_0}^2 } \right\} \, ,
\end{align}
which is expected to effectively express the quantum generation of the radiation. 

By using (\ref{fQG5}), one gets 
\begin{align}
\label{fQG5BBB}
f_1(Q) = \sqrt{-Q} \int^Q \frac{dq}{\sqrt{-q}} \left\{ - f_2\left( \mathcal{G} \left( q \right) \right) 
+ \mathcal{G} \left( q \right) f'_2 \left( \mathcal{G} \left( q \right) \right) 
+ \frac{4}{\sqrt{6}} \dot{\mathcal{G}} \left( q \right) f''_2 \left( \mathcal{G} \left( q \right) \right) \left( - q\right)^\frac{3}{2} 
 - \frac{q \left( q + 6 {H_0}^2 \right)}{2\kappa^2 \left( - q + 6 {H_0}^2 \right)} \right\}\, .
\end{align}
Here we use the expressions of $\mathcal{G}(q)$ and $\dot{\mathcal{G}}(q)$ by replacing $Q$ in (\ref{GQ1}) and (\ref{GQ2}) with $q$. 
Therefore, the explicit version of the theory which realises consistent inflation is found.

\subsubsection{Constant roll inflation}

Until now, there is not any confirmed observational evidence for the primordial non-Gaussianities. 
If the non-Gaussianities are found in future observations, a simple slow roll inflation model by a single scalar field might be excluded. 
As an alternative, the model of the constant-roll inflation model has been investigated. 
In the scalar-tensor theory, where a single scalar field $\phi$ couples with Einstein's gravity. the constant-roll condition is given by 
\begin{align}
\label{cnstr1}
\ddot\phi = \beta H \dot\phi\, ,
\end{align}
with a constant $\beta$. 
A solution under the condition (\ref{cnstr1}) is given by
\begin{align}
\label{cnstr2}
H=-M\tanh\left(\beta M t \right) \, ,
\end{align}
with a massive constant $M$. 
We now consider a model which reproduces (\ref{cnstr2}). 
Here we neglect the contribution from the matter because mainly matter could be generated after the inflation. 

Eq.~(\ref{cnstr2}) gives 
\begin{align}
\label{cnstr3}
Q =&\, - 6 M^2 \tanh^2\left(\beta M t \right) \, , \nonumber \\
\mathcal{G}
=&\, 
24 M^4 \tanh^2\left(\beta M t \right) \left( \tanh^2\left(\beta M t \right) + \frac{\beta}{\cosh^2 \left( \beta M t \right)} \right) 
= 
\frac{2}{3}Q \left\{ \left( 1 - \beta \right) Q - 6 \beta M^2 \right\}\, . \nonumber \\
\dot{\mathcal{G}}=&\, 
 - \frac{4\sqrt{6}}{3} M^2 \sqrt{-Q}\left\{ 2 \left( 1 - \beta \right) Q - 6 \beta M^2 \right\} \left( 1 + \frac{Q}{6M^2} \right)\, .
\end{align}
Then Eq.~(\ref{fQG5}) tells
\begin{align}
\label{fQG5cnstr1} 
f_1(Q) 
= \sqrt{-Q} \int^Q \frac{dq}{\sqrt{-q}} &\, \left\{ - f_2\left( \mathcal{G} \left( q \right) \right) 
+ \frac{2}{3}q \left\{ \left( 1 - \beta \right) q - 6 \beta M^2 \right\} f'_2 \left( \mathcal{G} \left( q \right) \right) \right. \nonumber \\
&\, \left. \quad - \frac{16}{3} M^2 q^2\left\{ 2 \left( 1 - \beta \right) q - 6 \beta M^2 \right\} \left( 1 + \frac{q}{6M^2} \right)
f''_2 \left( \mathcal{G} \left( q \right) \right) \right\}\, .
\end{align}
Especially when 
\begin{align}
\label{cnstr4}
f_2 \left( \mathcal{G} \right) = f_2^{(0)} {\mathcal{G}}^2\, ,
\end{align}
with a constant $f_2^{(0)}$, we obtain
\begin{align}
\label{fQG5cnstr2} 
f_1(Q) =
f_2^{(0)} \left\{ \frac{8}{81} \left( 1 - \beta \right) \left( 7 + \beta \right) \left( - Q\right) ^5 
+ \frac{32}{21} M^2 \left( 4 - 7 \beta + \beta^2 \right) \left( - Q \right)^4 
 - \frac{32}{5} M^4 \left( \beta^2 + 4 \beta \right) \left( - Q\right) ^3 + C\sqrt{-Q} 
\right\} \, .
\end{align}
Here $C$ is a constant of the integration. 
Hence we have shown that we obtain a model realising the constant-roll inflation. 

As in (\ref{Hex1b}), we may consider the transition from the constant-roll inflation to the radiation-dominated universe, 
instead of (\ref{cnstr2}), as follows, 
\begin{align}
\label{cnstr2rad}
H=-M\tanh\left(\frac{\beta M t - \sqrt{\beta^2 M^2 t^2 + 2 \beta}}{2}\right) \, ,
\end{align}
When $t$ is negative and large, $H$ behaves as (\ref{cnstr2}). 
On the other hand, when $t$ is positive and large, $H$ behaves as $H\sim \frac{1}{2t}$, which corresponds to the Hubble rate in the radiation-dominated universe. 
In the model (\ref{cnstr2rad}), the inflation may end at $t=0$. 

\subsection{Unification of inflation and dark energy epochs}

Let us now consider a model which describes both the inflation and dark energy epochs in a unified way. 
Such unification has been achieved in $f(R)$ gravity~\cite{Nojiri:2003ft, Nojiri:2010wj}. 
In this subsection, we use the $e$-folding number $N$ defined by $a=\e^N$ instead of the cosmological time. 

As proposed in \cite{Nojiri:2024zab}, we assume that the energy density $\rho$ is given by 
\begin{align}
\label{Uni1}
\rho(N) = \frac{\e^{n \left( N - N_0\right) }}{1 + \e^{n\left( N - N_0\right) }} \left( \rho_0^\mathrm{radiation} \e^{-4N} + \rho_0^\mathrm{baryon} \e^{-3N} \right) \, ,
\end{align}
where $\rho_0^\mathrm{radiation}$ and $\rho_0^\mathrm{baryon}$ are positive constants and $n$ is also a constant larger than $4$. 
We assume that the matter was generated at the end of the inflation as in (\ref{Hex1b3}). 
In addition to the radiation $\rho_0^\mathrm{radiation} \e^{-4N}$, 
we include the baryonic matter $rho_0^\mathrm{baryon} \e^{-3N}$. 
The factor $\frac{\e^{n\left( N-N_0\right) }}{1 + \e^{n\left( N-N_0\right) }}$ expresses the creation of matter 
and $N_0$ corresponds to the $e$-folding number when the inflation ends. 
When $N\ll N_0$, one finds $\frac{\e^{n\left( N-N_0\right) }}{1 + \e^{n\left(N-N_0\right) }} \sim \e^{n \left( N - N_0 \right)} \to 0$ and therefore $\rho(N)\to 0$, 
and when $N\gg N_0$, $\frac{\e^{N-N_0}}{1 + \e^{N-N_0}}\to 1$ and $\rho(N) \to \rho_0^\mathrm{radiation} \e^{-4N} + \rho_0^\mathrm{baryon} \e^{-3N}$. 

As in \cite{Nojiri:2024zab}, we consider the model that $Q$ and therefore $H$ are given by, 
\begin{align}
\label{Uni2}
Q =&\, -{ 6}H^2 = - \frac{ { 6}{H_0}^2 \left( 1 + \epsilon \e^{2 N - 2N_0} \right)}{1 + \e^{2 N - 2 N_0}} - { 2\kappa^2}\rho(N) \, ,
\end{align}
with positive constants $H_0$ and $\epsilon$. 
When $N\ll N_0$, the first term in the r.h.s. behaves as $- \frac{6{H_0}^2 \left( 1 + \epsilon \e^{N-N_0} \right)}{1 + \e^{N-N_0}} \to - { 6}{H_0}^2$ when $N\ll N_0$, 
which corresponds to the large effective cosmological constant generating inflation. 
On the other hand, when $N\gg N_0$, the first term behaves as $- \frac{ { 6}{H_0}^2 \left( 1 + \epsilon \e^{N-N_0} \right)}{1 + \e^{N-N_0}} \to -{ 6}\epsilon{H_0}^2$, 
which gives the small effective cosmological constant generating the late-time accelerating expansion by choosing $\epsilon$ to be very small. 

In principle, we obtain $N$ as a function of $Q$, $N=N(Q)$ by solving Eq.~(\ref{Uni2}). 
By substituting expression $N=N(Q)$ into (\ref{Uni1}), $\rho$ could be expressed as a function of $Q$, $\rho=\rho(Q)$. 
Then by using (\ref{fQG5}), we find the form of $f_1\left(Q\right)$. 

One may consider the unification of the constant-roll inflation and the dark energy. 
In the case of constant-roll inflation in (\ref{cnstr2}), the scale factor $a(t)$ is given by 
\begin{align}
\label{cnstr2a}
a=\e^{N-N_0} = \cosh^{-\frac{1}{\beta}} \left(\beta M t \right) \, ,
\end{align}
with a constant $N_0$. Therefore we find 
\begin{align}
\label{cnstr2N}
H= - M \sqrt{ 1 - \e^{\beta \left( N - N_0 \right)}} \, .
\end{align}
Note that when $t\to - \infty$, it follows $N\to -\infty$. 
Instead of (\ref{Uni2}), we propose the following, 
\begin{align}
\label{Uni2const}
Q =&\, - 6 H^2 = - 6 M^2 \left\{ 1 - \frac{\e^{\beta \left( N - N_0 \right)}}{1 + \left(1 + \epsilon^2 \right) \e^{\beta \left( N - N_0 \right)}} \right\}
 - { 2\kappa^2}\rho(N) \, ,
\end{align}
Here $\epsilon$ is a small and positive constant. 
When $N$ is negative and large, $ - 6 M^2 \left\{ 1 - \frac{\e^{\beta \left( N - N_0 \right)}}{1 + \left(1 + \epsilon^2 \right) \e^{\beta \left( N - N_0 \right)}} \right\}
 \to - 6 M^2 \left( 1 - \e^{\beta \left( N - N_0 \right)} \right)$, 
which corresponds to the constant-roll inflation in (\ref{cnstr2N}). 
On the other hand, when $N$ is positive and large, 
$ - 6 M^2 \left\{ 1 - \frac{\e^{\beta \left( N - N_0 \right)}}{1 + \left(1 + \epsilon^2 \right) \e^{\beta \left( N - N_0 \right)}} \right\} \to - 6 M^2 \epsilon^2$, 
which may correspond to the effective small cosmological constant generating the accelerating expansion of the present universe. 
Hence, the unified description of early-time inflation with late-time acceleration is constructed in the theory under consideration.

\subsection{Autonomous dynamical system} 

When one neglects the matter by putting $\rho=0$, by using (\ref{G00}) and the definition of the Gauss-Bonnet invariant 
$\mathcal{G}=24 H^2 \left( H^2 + \dot H \right)=- 4 Q \left( - \frac{Q}{6} + \dot H \right)$, we can rewrite the equations as 
an autonomous dynamical system: 
\begin{align}
\label{Atnm1}
\frac{d\mathcal{G}}{dN}=&\, \frac{3}{2Q^2 f_{\mathcal{G}\mathcal{G}}} \left( - f + 2 Q f_Q 
+ \mathcal{G}f_\mathcal{G} \right) \, , \\
\label{Atnm1B}
\frac{dQ}{dN}=&\, - 12 \left( - \frac{\mathcal{G}}{4Q} + \frac{Q}{6} \right) \, .
\end{align}
Here we have used the relation $\frac{d}{dt}= H\frac{d}{dN}$ between the derivatives with respect to the cosmological time $t$ and $e$-folding number $N$. 
For the dynamical analysis of $f(Q)$ gravity, see also \cite{Paliathanasis:2023nkb, Paliathanasis:2023raj}. 

For the autonomous dynamical system, it is often used dimensionless quantities but because we have not specified $f\left( Q, \mathcal{G} \right)$, we use the dimensional 
quantities like $Q$ and $\mathcal{G}$. 
The fixed point defined by $\frac{d\mathcal{G}}{dN}=\frac{dQ}{dN}=0$ corresponds to de Sitter spacetime, where the Hubble rate $H$ 
is a constant $H=H_0$. 
Then the conditions for the fixed point are, 
\begin{align}
\label{Atnm2}
0=- f + 2 Q f_Q 
+ \mathcal{G}f_\mathcal{G} \, , \quad 
0=- \frac{\mathcal{G}}{4Q} + \frac{Q}{6} \, .
\end{align}
The second equation (\ref{Atnm2}) gives a consistency that $\mathcal{G}=\frac{2}{3} Q^2=24 {H_0}^4$. 
Then the first equation has the following form, 
\begin{align}
\label{Atnm3}
0=- f\left( - 6 {H_0}^2, 24 {H_0}^4 \right) - 12 {H_0}^2 f_Q\left( - 6 {H_0}^2, 24 {H_0}^4 \right) 
+ 24 {H_0}^4 f_\mathcal{G} \left( - 6 {H_0}^2, 24 {H_0}^4 \right) \, .
\end{align}
If (\ref{Atnm3}) has a real and positive solution for ${H_0}^2$, there exists a solution describing the de Sitter spacetime. 
The autonomous equations in (\ref{Atnm1}) also show the stability of the de Sitter spacetime solution. 

We may also include matter whose EoS parameter is a constant $w$. 
Then the conservation law is written as $o=\dot\rho + 3H \left( 1 + w \right) \rho$, that is 
\begin{align}
\label{Atnm4}
\frac{d\rho}{dN}=-3 \left( 1 + w \right) \rho \, .
\end{align}
By including matter, Eq.~(\ref{Atnm1}) is modified as 
\begin{align}
\label{Atnm5}
\frac{d\mathcal{G}}{dN}= \frac{3}{2Q^2 f_{\mathcal{G}\mathcal{G}}} \left( - f + 2 Q f_Q 
+ \mathcal{G}f_\mathcal{G} + \rho \right) \, .
\end{align}
Therefore the corresponding autonomous dynamical system is given by Eqs.~(\ref{Atnm1B}), (\ref{Atnm4}), and (\ref{Atnm5}). 
The fixed point is given by the de Sitter spacetime, where the equations in (\ref{Atnm2}) are satisfied and $\rho=0$. 
The inclusion of matter may change the stability of the fixed point. 

As an example, we consider the following model, 
\begin{align}
\label{Atnm6}
f \left( Q, \mathcal{G} \right) = Q + f_1^{(0)} Q^2 + f_2^{(0)} {\mathcal{G}}^2\, ,
\end{align}
with constants, $f_1^{(0)}$ and $f_2^{(0)}$.
Then the fixed point is given by $\rho=0$ and (\ref{Atnm3}), which gives, 
\begin{align}
\label{Atnm7}
0=
 - 6 {H_0}^2 \left( 1 - 18 f_1^{(0)} {H_0}^2 + 96 f_2^{(0)} {H_0}^6 \right)\, .
\end{align}
In (\ref{Atnm7}), there is a trivial solution $H_0=0$ but because the denominator of (\ref{Atnm5}) also vanishes there, and therefore 
$\frac{d\mathcal{G}}{dN}$ diverges, 
\begin{align}
\label{Atnm5B}
\left.\frac{d\mathcal{G}}{dN}\right|_{H=H_0\to 0} \to \frac{1}{4 f_2^{(0)}{H_0}^2} \to \infty \, .
\end{align}
Therefore $H_0=0$ is not a fixed point. 
Besides a trivial solution $H_0=0$ in (\ref{Atnm7}), there may be non-trivial solutions
\begin{itemize}
\item When $f_1^{(0)}<0$ and $f_2^{(0)}>0$, there is no non-trivial solution besides $H_0=0$. 
\item When $f_1^{(0)}>0$ and $f_2^{(0)}<0$, there is one non-trivial solution besides $H_0=0$. 
\item When $f_1^{(0)}>0$ and $f_2^{(0)}>0$, 
\begin{itemize}
\item if $3 \sqrt{ \frac{3{f_1^{(0)}}^3}{f_2^{(0)}}}>1$, there are two non-trivial solutions besides $H_0=0$.
\item if $3 \sqrt{ \frac{3{f_1^{(0)}}^3}{f_2^{(0)}}}=1$, there is one non-trivial solution besides $H_0=0$.
\item if $3 \sqrt{ \frac{3{f_1^{(0)}}^3}{f_2^{(0)}}},1$, there is no non-trivial solution besides $H_0=0$.
\end{itemize}
\item When $f_1^{(0)}<0$ and $f_2^{(0)}<0$, there is one non-trivial solution besides $H_0=0$.
\end{itemize}
Therefore there appear non-trivial fixed points in general. 

We now investigate the stability of the fixed point(s). 
For this purpose, we consider the following perturbation, 
\begin{align}
\label{Atnm8}
Q=-6{H_0}^2 + \delta Q \, , \quad \mathcal{G}=- 24 {H_0}^4 + \delta\mathcal{G}\, , \quad 
\rho=\delta\rho\, .
\end{align}
Then Eqs.~(\ref{Atnm1B}), (\ref{Atnm4}), and (\ref{Atnm5}) give
\begin{align}
\label{Atnm5p}
\frac{d\delta\mathcal{G}}{dN}=&\, \frac{1}{48 f_2^{(0)} {H_0}^4} \left\{ 48 f_2^{(0)}{H_0}^4 \delta\mathcal{G} + \left( 1 - 36 f_1^{(0)} {H_0}^2 \right) \delta Q 
+ \delta\rho \right\} \, , \\
\label{Atnm1Bp}
\frac{d\delta Q}{dN}=&\, - \frac{1}{2{H_0}^2} \left( \delta\mathcal{G} + 8{H_0}^2 \delta Q \right) \, , \\
\label{Atnm4p}
\frac{d\delta\rho}{dN}=&\,-3 \left( 1 + w \right) \delta\rho \, . 
\end{align}
The eigenvalues of the matrix
\begin{align}
\label{Atnm9}
M
=\left( \begin{array}{ccc}
1 & \frac{1 - 36 f_1^{(0)} {H_0}^2 }{48 f_2^{(0)} {H_0}^4} & \frac{1}{48 f_2^{(0)} {H_0}^4} \\
 - \frac{1}{2{H_0}^2} & - 4 & 0 \\
0 & 0 & -3 \left( 1 + w \right) 
\end{array} \right) \, ,
\end{align}
are given by 
\begin{align}
\label{Atnm10}
\lambda_\pm \frac{- 3 \pm \sqrt{9 + \left( 16 - \frac{1 - 36 f_1^{(0)} {H_0}^2}{24 f_2^{(0)} {H_0}^6} \right)} }{2} \, , 
\lambda_3 = -3 \left( 1 + w \right) 
\end{align}
In order for the fixed point(s) to be stable, all the eigenvalues should be negative, which require
\begin{align}
\label{Atnm11}
\frac{1 - 36 f_1^{(0)} {H_0}^2}{24 f_2^{(0)} {H_0}^6} > 16\, , \quad 
w>-1 \, ,
\end{align}
The second condition tells that the phantom with $w<-1$ makes the fixed point unstable, as expected. 
Eq.~(\ref{Atnm7}) shows $0= 1 - 18 f_1^{(0)} {H_0}^2 + 96 f_2^{(0)} {H_0}^6$ and by eliminating $f_2^{(0)}$ from the first equation in (\ref{Atnm11}), 
we find 
\begin{align}
\label{Atnm12}
 - \frac{1 - 36 f_1^{(0)} {H_0}^2}{1 - 18 f_1^{(0)} {H_0}^2} > 4\, ,
\end{align}
which proves that for stability, we need to require 
\begin{align}
\label{Atnm13}
6> 108 f_1^{(0)} {H_0}^2 > 5\, .
\end{align}
Therefore the parameter $f_1^{(0)}$ should be positive. 
Following this direction, the autonomous dynamical system with more complicated matter content (including CDM or axion DM) may be studied.

\subsection{Gravitational wave}

In the case of the Einstein-Gauss-Bonnet theories, the models are constrained by the GW170817 neutron star merger event 
\cite{LIGOScientific:2017vwq, LIGOScientific:2017zic, LIGOScientific:2017ync}. 
The event indicates that the speed of the gravitational waves $c_\mathrm{GW}$ should nearly coincide with that of the light $c$ in the vacuum, 
\begin{align}
\label{GWp9} \left| \frac{{c_\mathrm{GW}}^2}{c^2} - 1 \right| < 6
\times 10^{-15}\, ,
\end{align}
This imposed a severe constraint on the form of the scalar coupling potential
for the Gauss-Bonnet invariant. 
Several realistic scenarios have been proposed for 
GW170817-compatible Einstein-Gauss-Bonnet theory
\cite{Odintsov:2020xji, Odintsov:2020sqy, Oikonomou:2021kql, Oikonomou:2022ksx}. 
The constraint used in these papers is valid only in the FLRW spacetime. 
Moreover, it cannot be satisfied around the static and spherically symmetric spacetime \cite{Nojiri:2023jtf}. 
In this section, we consider a similar constraint in the framework of $f \left( Q,\mathcal{G} \right)$ gravity. 

We now consider gravitational waves corresponding to the massless and spin-two modes although there could be a scalar ghost in $\mathcal{G}$ sector. 
For this purpose, one may take the perturbation of the metric $g_{\mu\nu} \to g_{\mu\nu} + h_{\mu\nu}$. 
For the modes which we are interested in, the following conditions should be satisfied, 
\begin{align}
\label{QGGW1}
h_{t\mu}=h_{\mu t}=0\, , \quad \sum_{i=1,2,3} h_{ii} = 0\, , \quad \nabla^i h_{ij}=\nabla^i h_{ji} = 0 \, .
\end{align}
Note that $Q$ and $\mathcal{G}$ are invariant under the variation $g_{\mu\nu} \to g_{\mu\nu} + h_{\mu\nu}$ with (\ref{QGGW1}). 
Although the full expression of the equation describing the gravitational wave is very complicated and lengthy, 
we are also now interested in the propagation speed of the gravitational wave. 
For the purpose, we only need to investigate the terms including the second-order derivative of $h_{\mu\nu}$. 
Then Eq.~(\ref{fQG2}) tells 
\begin{align}
\label{QGGW2}
0 \sim&\, \frac{1}{2} f_Q \partial^2 h_{\mu\nu} \nonumber \\
&\, - \frac{1}{2}\left\{ - 4 \partial^2 f_\mathcal{G} \delta^\tau_{\ \mu} \delta^\eta_{\ \nu}
+ 4 \left( \partial_\rho \partial_\mu f_\mathcal{G}\right) \delta^\eta_{\ \nu} g^{\rho\tau}
+ 4 \left( \partial_\rho \partial_\nu f_\mathcal{G}\right) \delta^\tau_{\ \mu} g^{\rho\eta}
 - 4g_{\mu\nu} \partial^\tau \partial^\eta f_\mathcal{G} \right\} \partial^2 h_{\tau\eta} \nonumber \\
&\, - 2 \left(\partial^\rho \partial^\sigma f_\mathcal{G} \right)
\left\{ \partial_\nu \partial_\rho h_{\sigma\mu}
 - \partial_\nu \partial_\mu h_{\sigma\rho}
 - \partial_\sigma \partial_\rho h_{\nu\mu}
 + \partial_\sigma \partial_\mu h_{\nu\rho} \right\} \, .
\end{align}
Here $\partial^2 \equiv \partial^\alpha \partial_\alpha$. 
Eq.~(\ref{QGGW2}) tells that if $\partial_\mu \partial_\nu f_\mathcal{G}$ is proportional to the metric $g_{\mu\nu}$,
\begin{align}
\label{condition}
\partial_\mu \partial_\nu f_\mathcal{G} = \frac{1}{4}g_{\mu\nu} \partial^2 f_\mathcal{G} \, ,
\end{align}
the speed of the gravitational wave is not changed from that of the light as found in \cite{Nojiri:2023jtf} for the
Einstein-Gauss-Bonnet gravity coupled with only one scalar field. 

In the FLRW spacetime, where $Q$ and $\mathcal{G}$ only depends on the cosmological time $t$, the condition (\ref{condition}) can be satisfied is 
$f_\mathcal{G}$ is a linear function of $t$, $f_\mathcal{G} = f_0 + f_1 t$. 
Here $f_0$ and $f_1$ are constants. 
The condition (\ref{condition}) cannot be, however, satisfied for the static and spherically symmetric spacetime as in the case of the Einstein-Gauss-Bonnet theories 
\cite{Nojiri:2023jtf}. 
Therefore the existence of the Gauss-Bonnet term makes the situation highly non-trivial and the term is constrained by the observations. 

There could be other kinds of fluctuations than gravitational waves of massless and spin two. 
Because the model includes the matter which is perfect fluid, there appears a sound wave, whose speed is given by 
${c_\mathrm{sound}}^2 = \frac{dp}{d\rho}= w$. 
Here $p$ and $\rho$ are pressure and the energy density of the matter and $w$ is the EoS parameter $p=w\rho$. 
In the case of the $f(R)$ gravity, there is a propagating scalar mode but in the case of the $f(Q)$ gravity, 
as shown in \cite{Capozziello:2024vix}, the only propagating mode in the flat background is a massless and spin-two mode corresponding to the gravitational wave 
and there are no other kinds of propagating modes like scalar mode. 
This may suggest that there could not be other propagating modes than the graviton although there is an open question of the ghost. 
Even in the modified gravity theories, we rewrite the FLRW equations in the form of 
$\frac{3H^2}{\kappa^2}=\rho_\mathrm{eff}$, $- \frac{1}{\kappa^2} \left( 2\dot H + 3H^2 \right)=p_\mathrm{eff}$. 
Here $\rho_\mathrm{eff}$ and $p_\mathrm{eff}$ are the effective energy density and the effective pressure. 
Then formally we may define the sound speed by 
${c_\mathrm{sound}}^2 = \frac{dp_\mathrm{eff}}{d\rho_\mathrm{eff}}= \frac{\frac{dp_\mathrm{eff}}{dt}}{\frac{d\rho_\mathrm{eff}}{dt}}$, for example. 
We should note, however, there is not the mode propagating with the speed of $c_\mathrm{eff}$ in the modified gravity theories. 

\section{$QR\mathcal{G}$ gravity\label{SecVII}}

As we mentioned, $f(Q, \tilde R)$ gravity is identical to $f(Q, B)$ in \cite{Capozziello:2023vne} or $f(Q, C)$ gravity in \cite{Gadbail:2023mvu}. 
So far, we have discussed $f $ gravity in detail. 
In this section, we briefly discuss a generalised example, that is, $f(Q, \tilde R, \mathcal{G})$ gravity in (\ref{Gnrl1QRG}) for completeness. 

There might look to be a special class of $f(Q, \tilde R, \mathcal{G})$ gravity, that is, the Einstein-$f(Q, \mathcal{G})$ gravity, whose action is given by 
\begin{align}
\label{EFQG}
S_{\mathrm{Einstein-}f(Q, \mathcal{G})} = \int d^4s \sqrt{-g} \left( \frac{\tilde R}{2\kappa^2} + f(Q, \mathcal{G}) \right)\, .
\end{align}
Because the difference between the scalar curvature $\tilde R$ and $Q$ is a total derivative, this model is equivalent to the $f(Q, \mathcal{G})$ gravity, 
where the redefined $f(Q, \mathcal{G})$ is given by $f_\mathrm{redefined} (Q, \mathcal{G})=\frac{Q}{2\kappa^2} + f(Q, \mathcal{G})$. 

The equation corresponding to (\ref{fQG2}) has the following form, 
\begin{align}
\label{fQRG}
0=&\, T_{\mu\nu} + \frac{1}{2} g_{\mu\nu} f 
 - f_Q g^{\alpha\beta} g^{\gamma\rho} \left\{ -\frac{1}{4} \nabla_\mu g_{\alpha\gamma} \nabla_\nu g_{\beta\rho}
 - \frac{1}{2} \nabla_\alpha g_{\mu\gamma} \nabla_\beta g_{\nu\rho} \right. \nonumber \\
&\, + \frac{1}{2} \left( \nabla_\mu g_{\alpha\gamma} \nabla_\rho g_{\beta\nu}
+ \nabla_\nu g_{\alpha\gamma} \nabla_\rho g_{\beta\mu} \right)
+ \frac{1}{2} \nabla_\alpha g_{\mu\gamma} \nabla_\rho g_{\nu\beta}
+ \frac{1}{4} \nabla_\mu g_{\alpha\beta} \nabla_\nu g_{\gamma\rho}
+ \frac{1}{2} \nabla_\alpha g_{\mu\nu} \nabla_\beta g_{\gamma\rho} \nonumber \\
&\, \left. - \frac{1}{4} \left( \nabla_\mu g_{\alpha\beta} \nabla_\gamma g_{\nu\rho} + \nabla_\nu g_{\alpha\beta} \nabla_\gamma g_{\mu\rho} \right)
 - \frac{1}{2} \nabla_\alpha g_{\mu\nu} \nabla_\gamma g_{\beta\rho}
 - \frac{1} {4} \left(\nabla_\alpha g_{\gamma\rho} \nabla_\mu g_{\beta\nu} + \nabla_\alpha g_{\gamma\rho} \nabla_\nu g_{\beta\mu} \right)
\right\} \nonumber \\
&\, - \frac{g_{\mu\rho} g_{\nu\sigma}}{\sqrt{-g}} \partial_\alpha \left[ \sqrt{-g} f_Q \left\{ - \frac{1}{2} g^{\alpha\beta} g^{\gamma\rho} g^{\sigma\tau} \nabla_\beta g_{\gamma\tau}
+ \frac{1}{2} g^{\alpha\beta} g^{\gamma\rho} g^{\sigma\tau} \left( \nabla_\tau g_{\gamma\beta} + \nabla_\gamma g_{\tau\beta} \right) \right. \right. \nonumber \\
&\, \left. \left. + \frac{1}{2} g^{\alpha\beta} g^{\rho\sigma} g^{\gamma\tau} \nabla_\beta g_{\gamma\tau}
 - \frac{1}{2} g^{\alpha\beta} g^{\rho\sigma} g^{\gamma\tau} \nabla_\gamma g_{\beta\tau}
 - \frac{1}{4} \left( g^{\alpha\sigma} g^{\gamma\tau} g^{\beta\rho} + g^{\alpha\rho} g^{\gamma\tau} g^{\beta\sigma}
\right) \nabla_\beta g_{\gamma\tau}
\right\} \right] \nonumber \\
&\, - \frac{g_{\mu\rho} g_{\nu\sigma}}{\sqrt{-g}} {\Gamma^\rho}_{\alpha\eta} \left[ \sqrt{-g} f_Q 
\left\{ - \frac{1}{2} g^{\alpha\beta} g^{\gamma\eta} g^{\sigma\tau} \nabla_\beta g_{\gamma\tau}
+ \frac{1}{2} g^{\alpha\beta} g^{\gamma\eta} g^{\sigma\tau} \left( \nabla_\tau g_{\gamma\beta} + \nabla_\gamma g_{\tau\beta} \right) \right. \right. \nonumber \\
&\, \left. \left. + \frac{1}{2} g^{\alpha\beta} g^{\eta\sigma} g^{\gamma\tau} \nabla_\beta g_{\gamma\tau}
 - \frac{1}{2} g^{\alpha\beta} g^{\eta\sigma} g^{\gamma\tau} \nabla_\gamma g_{\beta\tau}
 - \frac{1}{4} \left( g^{\alpha\sigma} g^{\gamma\tau} g^{\beta\eta} + g^{\alpha\eta} g^{\gamma\tau} g^{\beta\sigma}
\right) \nabla_\beta g_{\gamma\tau}
\right\} \right] \nonumber \\
&\, - \frac{g_{\mu\rho} g_{\nu\sigma}}{\sqrt{-g}} {\Gamma^\sigma}_{\alpha\eta} \left[ \sqrt{-g} f_Q 
\left\{ - \frac{1}{2} g^{\alpha\beta} g^{\gamma\rho} g^{\eta\tau} \nabla_\beta g_{\gamma\tau}
+ \frac{1}{2} g^{\alpha\beta} g^{\gamma\rho} g^{\eta\tau} \left( \nabla_\tau g_{\gamma\beta} + \nabla_\gamma g_{\tau\beta} \right) \right. \right. \nonumber \\
&\, \left. \left. + \frac{1}{2} g^{\alpha\beta} g^{\rho\eta} g^{\gamma\tau} \nabla_\beta g_{\gamma\tau}
 - \frac{1}{2} g^{\alpha\beta} g^{\rho\eta} g^{\gamma\tau} \nabla_\gamma g_{\beta\tau}
 - \frac{1}{4} \left( g^{\alpha\eta} g^{\gamma\tau} g^{\beta\rho} + g^{\alpha\rho} g^{\gamma\tau} g^{\beta\eta}
\right) \nabla_\beta g_{\gamma\tau}
\right\} \right] \nonumber \\
&\, - \tilde R_{\mu\nu}f_{\tilde R} + \tilde\nabla_\mu \tilde\nabla_\nu f_{\tilde R} - g_{\mu\nu} \tilde\nabla^2 f_{\tilde R} 
 -2 f_\mathcal{G} {\tilde R} {\tilde R}_{\mu\nu}
+ 4f_\mathcal{G} {\tilde R}_{\mu\rho} {\tilde R}_\nu^{\ \rho} 
 -2 f_\mathcal{G} {\tilde R}_\mu^{\ \rho\sigma\tau} {\tilde R}_{\nu\rho\sigma\tau} \nonumber \\
&\, - 4 f_\mathcal{G} {\tilde R}_{\mu\rho\sigma\nu} {\tilde R}^{\rho\sigma} 
+ 2 \left( \tilde\nabla_\mu \tilde\nabla_\nu f_\mathcal{G} \right){\tilde R} 
 - 2 g_{\mu\nu} \left( \tilde\nabla^2 f_\mathcal{G} \right) {\tilde R} 
 - 4 \left( \tilde\nabla^\rho \tilde\nabla_\mu f_\mathcal{G} \right) {\tilde R}_{\nu\rho} 
 - 4 \left( \tilde\nabla^\rho \tilde\nabla_\nu f_\mathcal{G} \right){\tilde R}_{\mu\rho} \nonumber \\
&\, + 4 \left( \tilde\nabla^2 f_\mathcal{G} \right){\tilde R}_{\mu\nu} 
+ 4g_{\mu\nu} \left( \tilde\nabla_{\rho} \tilde\nabla_\sigma f_\mathcal{G} \right) {\tilde R}^{\rho\sigma}
 - 4 \left(\tilde\nabla^\rho \tilde\nabla^\sigma f_\mathcal{G} \right) {\tilde R}_{\mu\rho\nu\sigma} \, .
\end{align}
Because this equation includes the fourth derivatives of the metric as in the standard $f(R)$ gravity, we expect that there could appear a propagating scalar mode. 
In the case of $f(R)$ gravity, the action can be rewritten in the scalar-tensor form and the scalar field corresponds to the scale of the metric. 
In $f(Q)$ gravity, however, the corresponding scalar mode becomes a ghost but due to the constraint, it does not propagate \cite{Hu:2023gui}. 
Therefore we need a more detailed investigation by using the first class and the second class constraints in the framework of the Hamiltonian analysis. 

In principle, one may consider a more general case. 
Because any scalar quantity including curvature is given by a combination of the metric $g_{\mu\nu}$ and the Riemann curvature $R_{\mu\nu\rho\sigma}$, 
more general action may be given by 
\begin{align}
\label{Gnrl1}
S_\mathrm{general} = \int d^4x \sqrt{-g} f_\mathrm{general} \left(Q, g_{\mu\nu}, {\tilde R}_{\mu\nu\rho\sigma} \right)\, .
\end{align}
Here $g_{\mu\nu}$ expresses the metric which is not included in $Q$ and ${\tilde R}_{\mu\nu\rho\sigma}$ in the arguments. 
The equation corresponding to (\ref{fQRG}) can be easily found by using the following formula of the variation with respect to the metric, 
\begin{align}
\label{deltariemann}
\delta \tilde R_{\mu\nu\rho\sigma}=&\, \frac{1}{2}\left[\tilde\nabla_\rho \tilde\nabla_\nu \delta g_{\sigma\mu}
 - \tilde\nabla_\rho \tilde\nabla_\mu \delta g_{\sigma\nu}
 - \tilde\nabla_\sigma \tilde\nabla_\nu \delta g_{\rho\mu}
 + \tilde\nabla_\sigma \tilde\nabla_\mu \delta g_{\rho\nu}
+ \delta g_{\mu\tau} \tilde R^\tau_{\ \nu\rho\sigma}
 - \delta g_{\nu\tau} \tilde R^\tau_{\ \mu\rho\sigma} \right] \, .
\end{align}
Therefore we may obtain the equations derived from the general action (\ref{Gnrl1}). 

Of course, $f\left(Q, \mathcal{G} \right)$ gravity and $f\left(Q, {\tilde R}, \mathcal{G} \right)$ gravity are special cases of the general model (\ref{Gnrl1}). 
For example, $f\left(Q, R, \mathcal{G} \right)$ corresponds to the following $f_\mathrm{general} \left(Q, g_{\mu\nu}, {\tilde R}_{\mu\nu\rho\sigma} \right)$, 
\begin{align}
\label{general2}
f_\mathrm{general} &\, \left(Q, g_{\mu\nu}, {\tilde R}_{\mu\nu\rho\sigma} \right) \nonumber \\
=&\, f\left(Q, {\tilde R}=g^{\mu\nu} g^{\rho\sigma} {\tilde R}_{\mu\rho\nu\sigma}, \nonumber \right. \\
&\, \mathcal{G}=\frac{1}{4}\left( 
g^{\alpha\mu} g^{\beta\nu} g^{\gamma\rho} g^{\delta\sigma} 
 - g^{\alpha\mu} g^{\beta\nu} g^{\gamma\sigma} g^{\delta\rho}
+ g^{\alpha\mu} g^{\beta\rho} g^{\gamma\sigma} g^{\delta\nu} 
 - g^{\alpha\mu} g^{\beta\rho} g^{\gamma\nu} g^{\delta\sigma} 
+ g^{\alpha\mu} g^{\beta\sigma} g^{\gamma\nu} g^{\delta\rho} 
 - g^{\alpha\mu} g^{\beta\sigma} g^{\gamma\rho} g^{\delta\nu} \right. \nonumber \\
& \qquad - g^{\alpha\nu} g^{\beta\rho} g^{\gamma\sigma} g^{\delta\mu}
+ g^{\alpha\nu} g^{\beta\rho} g^{\gamma\mu} g^{\delta\sigma}
 - g^{\alpha\nu} g^{\beta\sigma} g^{\gamma\mu} g^{\delta\rho}
+ g^{\alpha\nu} g^{\beta\sigma} g^{\gamma\rho} g^{\delta\mu}
 - g^{\alpha\nu} g^{\beta\mu} g^{\gamma\rho} g^{\delta\sigma}
+ g^{\alpha\nu} g^{\beta\mu} g^{\gamma\sigma} g^{\delta\rho} \nonumber \\
& \qquad + g^{\alpha\rho} g^{\beta\sigma} g^{\gamma\mu} g^{\delta\nu} 
 - g^{\alpha\rho} g^{\beta\sigma} g^{\gamma\nu} g^{\delta\mu} 
+ g^{\alpha\rho} g^{\beta\mu} g^{\gamma\nu} g^{\delta\sigma} 
 - g^{\alpha\rho} g^{\beta\mu} g^{\gamma\sigma} g^{\delta\nu} 
+ g^{\alpha\rho} g^{\beta\nu} g^{\gamma\sigma} g^{\delta\mu} 
 - g^{\alpha\rho} g^{\beta\nu} g^{\gamma\mu} g^{\delta\sigma} \nonumber \\
& \qquad \left. - g^{\alpha\sigma} g^{\beta\mu} g^{\gamma\nu} g^{\delta\rho} 
+ g^{\alpha\sigma} g^{\beta\mu} g^{\gamma\rho} g^{\delta\nu} 
 - g^{\alpha\sigma} g^{\beta\nu} g^{\gamma\rho} g^{\delta\mu} 
+ g^{\alpha\sigma} g^{\beta\nu} g^{\gamma\mu} g^{\delta\rho} 
 - g^{\alpha\sigma} g^{\beta\rho} g^{\gamma\mu} g^{\delta\nu} 
+ g^{\alpha\sigma} g^{\beta\rho} g^{\gamma\nu} g^{\delta\mu} \right) \nonumber \\
& \qquad \qquad \left. \times {\tilde R}_{\alpha\beta\gamma\delta} {\tilde R}_{\mu\nu\rho\sigma} \right) \, .
\end{align}
If the vierbein field is used, we may also include the Chern-Simons invariant. Therefore, following to above strategy one can propose and study $F(Q)$ gravity with higher-order Riemann invariants up to the necessary order.

\section{Summary and Discussion\label{SecVIII}}

The main result of this paper is that $f(Q)$ gravity or the symmetric teleparallel theory does not exclude the curvature in Einstein's gravity, that is, 
the Riemann curvature constructed from the standard Levi-Civita connection. 
The first point which we notice is that the conservation law of matter is usually not given by the covariant derivative of the symmetric teleparallel theory 
but the covariant derivative constructed from the Levi-Civita connection.
The commutator of the covariant derivative given by the Levi-Civita connection inevitably induces the Riemann curvature in Einstein's gravity. 
Symmetry or any fundamental principle does not exclude the curvature in Einstein's gravity and therefore they will also appear in the quantum corrections. 

This idea to include the curvatures in Einstein's gravity into $f(Q)$ gravity is not new and in \cite{Gadbail:2023mvu}, a model including $Q$ and Einstein's scalar curvature $\tilde R$ 
has been substantially proposed. 
This is one of the reasons why we concentrated on $f\left( Q, \mathcal{G} \right)$ gravity with the Gauss-Bonnet invariant $\mathcal{G}$. 
The cosmological applications of such a model are developed and the cosmological reconstruction is done. 
In the formulation of the reconstruction, we give a systematic way to construct a model which realises any given geometry. 

The main results of this paper are the following:
\begin{itemize}
\item We have shown that the inclusion of the curvatures in Einstein's gravity in $f(Q)$ gravity theory is not prohibited and might be natural. 
\item We gave some speculation about the ghost problem. 
\item By using the expressions of the field equations in the FLRW spacetime, we developed the formalism of the cosmological reconstruction. 
\item We applied the obtained formalism of the reconstruction and constructed or outlined the construction of explicit models, which realise, 
1) mimicking the $\Lambda$CDM model, 2) slow-roll inflation, 3) constant-roll inflation, 4) unification of inflation and dark energy epochs. 
In addition, the dynamical autonomous system in the theory under investigation is formulated. 
Some remarks on the gravitational wave in such a theory are made. 
\item We also generalized the model to $f\left( Q, \tilde R, \mathcal{G} \right)$ gravity and further generalised model. 
\end{itemize}

Maybe a critical problem could be ghosts. 
In the case of the Einstein-$f\left( \mathcal{G} \right)$ gravity, it is known that there appear ghosts but the elimination of the ghosts has been well discussed~\cite{Nojiri:2018ouv}. 
Even in the model under discussion, the ghosts originated from the $\mathcal{G}$ sector could be eliminated similarly but at present, it is not clear if there appear ghosts from the $Q$ sector. 
Therefore even if the ghosts exist, we do not know how we can eliminate them. 
We should note that there is the problem of the dynamical degrees of freedom (DOF) of $f(Q)$ gravity~\cite{Hu:2022anq, DAmbrosio:2023asf, Heisenberg:2023lru, 
Paliathanasis:2023pqp, Dimakis:2021gby, Hu:2023gui}. 
Although only propagating mode in the flat background could be a graviton \cite{Capozziello:2024vix}, as in the $f(T)$ gravity \cite{Bamba:2013ooa}, 
non-physical modes appear in the $f(T)$ gravity~\cite{Ong:2013qja, Izumi:2012qj}. 
Even in the case of the $f(Q)$ gravity, there might be a similar mode in the higher-order perturbation. 
We expect, however, that such unphysical modes could be eliminated by using the constraint as in (\ref{FRGBg20}). 
For this purpose, first, one must check the dynamical degrees of freedom. 
If any unphysical modes are found, they should be specified. 
There are several ways to eliminate the unphysical modes. 
One way is to construct a model so that the variations of the modes become gauge symmetry. 
Then we may fix the gauge for the modes to vanish. This will be considered elsewhere.
Another way could be to use the constraint as in (\ref{FRGBg20}). 
Anyway, we have proposed a new interesting class of non-metricity gravity models, whose consistency and comparison with  observational bounds 
may be checked in more detail in the near future.

\section*{ACKNOWLEDGEMENTS}

This work was partially supported by MICINN (Spain), project PID2019-104397GB-I00 and the program Unidad
de Excelencia Maria de Maeztu CEX2020-001058-M, Spain (S.D.O).

\end{document}